\documentclass[prd,aps,twocolumn,a4paper,floatfix,nofootinbib]{revtex4-2}

\usepackage[utf8]{inputenc}
\usepackage{graphicx,psfrag}
\usepackage{mathrsfs}
\usepackage{amsmath,amsfonts,amssymb}
\usepackage{multirow}
\usepackage{diagbox}
\usepackage{comment}
\usepackage{xcolor}
\usepackage{enumerate}
\usepackage{booktabs}
\usepackage[normalem]{ulem}
\usepackage{hyperref}
\usepackage{mathtools}
\usepackage{siunitx}
\usepackage{textgreek}
\hypersetup{
    colorlinks = true,
    linkcolor = {blue},
    citecolor = {blue},
    urlcolor = {blue},
    linkbordercolor = {white},
    }

\usepackage{color}
\definecolor{cyan}{rgb}{0,0.9,0.9}
\definecolor{orange}{rgb}{0.9,0.5,0}
\definecolor{magenta}{rgb}{1,0,1}
\definecolor{purple}{rgb}{0.8,0.4,0.8}
\definecolor{gray}{rgb}{0.8242,0.8242,0.8242}
\definecolor{green}{rgb}{0.,0.8,0.}

\def\bam{{\textsc{bam}}}
\def\spritz{{\textsc{spritz}}}
\def\sacra{{\textsc{sacra}}$_{\rm KK22}$}
\def\gramx{{\textsc{GRaM-X}}}

\usepackage[normalem]{ulem}

\begin{document}

\title{General Relativistic Magneto-Hydrodynamic Simulations with BAM: Implementation and Code Comparison}

\author{Anna \surname{Neuweiler}$^{1}$}
\author{Tim \surname{Dietrich}$^{1,2}$}
\author{Bernd \surname{Br\"ugmann}$^{3}$}
\author{Edoardo \surname{Giangrandi}$^{1,7}$}
\author{Kenta \surname{Kiuchi}$^{2,9}$}
\author{Federico \surname{Schianchi}$^{1}$}
\author{Philipp \surname{M\"osta}$^{8}$}
\author{Swapnil \surname{Shankar}$^{8}$}
\author{Bruno \surname{Giacomazzo}$^{4,5,6}$}
\author{Masaru \surname{Shibata}$^{2,9}$}

\affiliation{${}^1$ Institut f\"ur Physik und Astronomie, Universit\"at Potsdam, Haus 28, Karl-Liebknecht-Str. 24/25, 14476, Potsdam, Germany}
\affiliation{${}^2$ Max Planck Institute for Gravitational Physics (Albert Einstein Institute), Am M\"uhlenberg 1, Potsdam 14476, Germany}
\affiliation{${}^3$ Theoretical Physics Institute, University of Jena, 07743 Jena, Germany}
\affiliation{${}^4$ Dipartimento di Fisica G. Occhialini, Universit\`a di Milano-Bicocca, Piazza della Scienza 3, I-20126 Milano, Italy}

\affiliation{${}^5$ INFN, Sezione di Milano-Bicocca, Piazza della Scienza 3, I-20126 Milano, Italy}
\affiliation{${}^6$ INAF, Osservatorio Astronomico di Brera, Via E. Bianchi 46, I-23807 Merate, Italy}
\affiliation{${}^7$ CFisUC, Department of Physics, University of Coimbra, Rua Larga P-3004-516, Coimbra, Portugal}
\affiliation{${}^8$ GRAPPA, Anton Pannekoek Institute for Astronomy, Institute of High-Energy Physics, and Insitute of Theoretical Physics, University of Amsterdam, Science Park 904, 1098 XH Amsterdam, The Netherlands}
\affiliation{${}^9$ Center for Gravitational Physics and Quantum Information, Yukawa Institute for Theoretical Physics, Kyoto University, Kyoto, 606-8502, Japan}

\date{\today}

\begin{abstract}
Binary neutron star mergers are among the most energetic events in our Universe, with magnetic fields significantly impacting their dynamics, particularly after the merger. While numerical-relativity simulations that correctly describe the physics are essential to model their rich phenomenology, the inclusion of magnetic fields is crucial for realistic simulations. For this reason, we have extended the \bam\ code to enable general relativistic magneto-hydrodynamic (GRMHD) simulations employing a hyperbolic `divergence cleaning' scheme. We present a large set of standard GRMHD tests and compare the \bam\ code to other GRMHD codes, \spritz, \gramx, and \sacra, which employ different schemes for the evolution of the magnetic fields.  Overall, we find that the \bam\ code shows a good performance in simple special-relativistic tests. In addition, we find good agreement and consistent results when comparing GRMHD simulation results between \bam\ and \sacra. 
\end{abstract}

\maketitle

\section{Introduction}
\label{sec:Intro}

Magnetic fields play a crucial role in various astrophysical high-energy scenarios. In particular, they are important in binary neutron star (BNS) mergers because the magnetohydrodynamical effects can influence the lifetime of the merger remnant and shape the matter outflow in the post-merger phase. In the multi-messenger event GW170817~\cite{LIGOScientific:2017vwq}, the first observation of a BNS system by gravitational waves (GWs) accompanied by electromagnetic (EM) counterparts including the gamma-ray burst (GRB) GRB170817A~\cite{Goldstein:2017mmi,Savchenko:2017ffs}, strong evidence has been found for BNS merger to be sources of GRBs triggered by a relativistic jet, e.g., \cite{Troja:2017nqp,Hallinan:2017woc,Mooley:2017enz,Lazzati:2017zsj}. The launch of such a relativistic jet is most likely powered by magneto-hydrodynamic processes, e.g., \cite{Rezzolla:2011da,Ruiz:2016rai,Mosta:2020hlh,Kiuchi:2023obe}. \par 

A key in the study of BNS systems are numerical-relativity (NR) simulations that solve Einstein's field equations together with magneto-hydrodynamics equations. Due to the high complexity and dynamics of the merger process, an accurate description of the spacetime and hydrodynamics is crucial to allow for realistic predictions and to enable us to study details of the physical processes. Furthermore, we can extract the GW signal and the matter outflow from NR simulations to inform GW models as well as models for the EM counterparts, which are fundamental for the correct interpretation of multi-messenger observations. Among others, this allows for studies on the behavior of matter at supranuclear densities, e.g.,~\cite{Radice:2016rys,Margalit:2017dij,LIGOScientific:2018cki,Pang:2022rzc}, the expansion rate of the Universe, e.g.,~\cite{LIGOScientific:2017adf,LIGOScientific:2019zcs,Hotokezaka:2018dfi,LIGOScientific:2018gmd,Dietrich:2020efo}, and the synthesis of heavy elements, e.g.,~\cite{Lattimer:1974slx,Rosswog:1998hy,Korobkin:2012uy,Wanajo:2014wha}. In the last decade, there has been rapid progress towards more realistic representations of microphysics in NR simulations, such as using tabulated equations of state (EOS) based on nuclear physics calculations to describe the interior of neutron stars (NSs), including neutrino emission and transport, as well as including magnetic fields.\par 

In this work, we focus on the latter, i.e., the implementation of general relativistic magneto-hydrodynamics (GRMHD) routines in our NR code \bam~\cite{Bruegmann:2006ulg,Thierfelder:2011yi}, but also refer to previous works concentrating on employing tabulated EOSs~\cite{Gieg:2022mut} and on the implementation of an advanced multipolar first-order (M1) neutrino transfer scheme~\cite{Schianchi:2023uky} to highlight recent progress in performing more accurate simulations with \bam. Magnetic fields, although negligible for the inspiral, are amplified during and after the merger due to the Kelvin-Helmholz instability (KHI), e.g.,~\cite{Kiuchi:2015sga, Kiuchi:2017zzg,Giacomazzo:2014qba}, the Rayleigh-Taylor instability, e.g.,~\cite{Skoutnev:2021chg}, and the magneto-rotational instability (MRI), e.g.,~\cite{Siegel:2013nrw,Kiuchi:2023obe}. Accordingly, they influence in particular the evolution of the remnant system and the outflow of matter in the post-merger phase due to magnetically driven winds and jet formation, making their incorporation crucial for correct ejecta models and predictions of EM signals, e.g., \cite{Ciolfi:2020cpf}. At the same time, they significantly increase the complexity of the simulations and pose new numerical challenges. \par

One challenge in GRMHD is to fulfill the divergence-free condition for the magnetic field. There are various approaches in the literature to ensure that no magnetic monopoles are formed. Among the most popular schemes used in the NR community nowadays are the \textit{constraint transport} scheme as initially developed by \cite{Evans:1988}, e.g., \cite{Shankar:2022ful, Kiuchi:2022ubj, Cook:2023bag}, \textit{vector potential} methods, e.g. \cite{Etienne:2010ui, Cipolletta:2019geh}, and \textit{divergence cleaning} approaches, e.g.,\cite{Dedner:2002,Liebling:2010bn,Penner:2010px,Mosta:2013gwu, Deppe:2021bhi,Palenzuela:2018sly}. \textit{Constraint transport} uses the induction equation and Stokes theorem to integrate the magnetic field (see \cite{Toth:2000} for a detailed discussion of \textit{constraint transport} methods). In \textit{vector potential} methods, one exploits the fact that the magnetic field can be expressed by a \textit{vector potential}, which is then evolved instead. For both methods, the divergence of the magnetic field is by construction zero at round-off accuracy. However, both methods require a staggered grid, as the magnetic field components are usually defined on the surface of a grid cell, which complicates implementation in codes using mesh refinement. Also, \textit{vector potential} methods do not guarantee magnetic flux conservation across the refinement boundary when mesh refinement is employed. \par

We apply in \bam\ the hyperbolic \textit{divergence cleaning} approach following \cite{Liebling:2010bn,Mosta:2013gwu}. In this approach, a new field variable is introduced to damp and advect the divergence of the magnetic field. Hyperbolic \textit{divergence cleaning} of this type originated in \cite{Dedner:2002} and was introduced to NR in \cite{Liebling:2010bn} for magnetized rigidly rotating NSs. \cite{Mosta:2013gwu} showed very promising results for standard tests of special relativistic MHD, while \textit{divergence cleaning} encountered some preliminary issues for GRMHD, so their focus was on \textit{constraint transport} for GRMHD. 

\textit{Divergence cleaning} offers the advantage of a relatively simple implementation, but has limitations in maintaining an exact zero divergence of the magnetic field. In fact, we expect to find an order of $10^{-3}$ for the divergence of the magnetic field normalized to the magnetic field strength \cite{Palenzuela:2021gdo}. On the other hand, \textit{divergence cleaning} is directly compatible with high-order schemes for fluxes, and thus, high-resolution shock capturing schemes can be applied directly without large code changes. In codes with \textit{constraint transport} or \textit{vector potential}, the magnetic field needs to be interpolated from the cell surfaces to the cell center, which is usually done linearly. 

In the present work, the goal is to analyse how well \textit{divergence cleaning} actually performs compared to the other methods. For this reason, we conduct a comprehensive comparative analysis of our implementation with other, well-established GRMHD codes using different treatments for the divergence-free constraint: the \gramx\ code~\cite{Shankar:2022ful}, the \textsc{sacra} variant of \citet{Kiuchi:2022ubj} henceforth called \sacra, and the \spritz\ code~\cite{Cipolletta:2019geh}. The latter uses a \textit{vector potential} method, while the other two codes use \textit{constraint transport}. Such direct code comparisons are still rare, but are crucial to identify how dependent results are on the method used and how serious errors due to small divergences of the magnetic field are in the evolution also in relation to other numerical inaccuracies. In \cite{Espino:2022mtb}, a comparison for different open-source GRMHD codes showed discrepancies in BNS merger scenarios, particularly in the merger times and remnant lifetimes. \par 

Our code comparison includes relativistic shock tests in one, two and three dimensions as well as BNS merger simulations. While the basic structure of the codes is relatively similar, we have to consider in the comparison not only differences in the magnetic field treatment but also in several other numerical implementations, such as high-resolution shock capturing (HRSC) schemes or algorithms for flux conservation at refinement boundaries. By using higher and lower orders for the reconstruction methods of fluid variables and different grid setups, we can estimate their effects and try to separate them from differences caused by the divergence-free constraint treatments. In fact, the tests show that the latter are negligible compared to the effects caused by using different reconstruction methods.
For the BNS merger simulation, we run the setup also at three different resolutions to carefully analyze the resolution effects and to determine whether the differences become larger or smaller at higher resolutions. \par

The article is structured as follows: Section~\ref{sec:Methods} summarizes the relevant GRMHD evolutions equations, describes the divergence-free constraint treatment, and new numerical methods implemented in the \bam\ code. In Sec.~\ref{sec:FirstTests}, we present a series of relativistic tests performed with \bam\ and compare them with results of the GRMHD codes \gramx, \sacra, and \spritz. We then discuss results of BNS merger simulations performed with \bam\ and \sacra, and compare them in Sec.~\ref{sec:BNSSimulations}. A summary of our main results follows in Sec.~\ref{sec:Conclusions}. In this article, we apply a metric with $(-,+,+,+)$ signature and geometric units with $G=c=M_\odot=1$, unless otherwise specified. 

\section{Relevant Equations and Numerical Methods}
\label{sec:Methods}

We summarize in this section our new GRMHD implementation in the \bam\ code~\cite{Bruegmann:2006ulg,Thierfelder:2011yi,Dietrich:2015iva,Bernuzzi:2016pie}. Below, we briefly outline the framework of the NR code, including spacetime evolution and grid structure. We then discuss the relevant evolution equations and applied divergence-free constraint treatment. Further, we highlight new numerical techniques that we implemented in \bam\ to address the increased complexity of the systems due to magnetic fields. 

\subsection{Spacetime Evolution and Grid Structure}

\bam\ evolves the gravitational field in time using the methods-of-line approach and employing finite difference stencils for spatial discretization. For this, Einstein's field equations are solved in a 3+1 decomposition, in which the line element is:
\begin{equation}
    ds^2 = -\alpha^2dt^2 +  \gamma_{ij}(dx^i + \beta^i dt)(dx^j + \beta^j dt),
\end{equation}
with $\alpha$ as lapse function, $\beta^i$ as shift vector, and $\gamma_{ij}$ as spatial part of the metric tensor $g_{\mu\nu}$ induced on a three-dimensional, spatial hypersurface. The latter is given by:
\begin{equation}
    \gamma_{\alpha \beta} = g_{\alpha\beta} + n_{\alpha}n_{\beta},
\end{equation}
where $n^{\alpha}$ is the timelike, normal vector to the three-dimensional hypersurface defined by: 
\begin{equation}
    n^\mu = \frac{1}{\alpha} \left(1, -\beta^i\right) , \hspace{1cm} n_\mu = \left(-\alpha,0,0,0\right).
\end{equation}

In this work, we apply a fourth-order Runge-Kutta integration and a Courant-Friedrichs-Lewy (CFL) coefficient of 0.25 in all \bam\ simulations. We use the BSSN reformulation \cite{Nakamura:1987zz,Shibata:1995we,Baumgarte:1998te} for the tests presented in Sec.~\ref{sec:FirstTests} and the Z4c reformulation with constraint damping terms \cite{Bernuzzi:2009ex,Hilditch:2012fp} for the BNS simulations presented in Sec.~\ref{sec:BNSSimulations}, combined with 1+log slicing~\cite{Bona:1994a} and gamma-driver shift conditions~\cite{Alcubierre:2002kk}. \\

The infrastructure of \bam\ includes adaptive mesh refinement (AMR). The grid consists of a hierarchy of $L$ refinement levels labeled by $l=0,...,L-1$ with one or more Cartesian boxes. Using a 2:1 refinement strategy, the grid spacing on each level is given by $h_l = h_0/2^l$. For inner refinement levels with $l \geq l_{\rm mv}$, the Cartesian boxes can move and adjust dynamically during the evolution to track the compact objects. Thereby, each box consists of $n$ grid points per direction or respectively $n_{\rm mv}$ grid points per direction for inner, moving boxes. The points are cell-centered and staggered to prevent division by zero at the origin. As a result, the grid points of the successive levels $l$ and $l+1$ do not coincide. \par

Furthermore, we apply in \bam\ the Berger-Oliger scheme~\citep{Berger:1984zza} for local-time stepping (see \cite{Bruegmann:2006ulg}) and the Berger-Colella scheme~\citep{Berger:1989a} to ensures flux conservation across refinement boundaries (see \cite{Dietrich:2015iva}).

\subsection{General Relativistic Magneto-Hydrodynamics}
\label{subsec:GRMHD}

We use the Valencia formulation of the GRMHD equations~\cite{Font:2008fka,Marti:1991wi,Banyuls:1997zz,Anton:2005gi}. We note that we apply the ideal GRMHD approximation assuming infinite conductivity and zero resistivity. Although a resistive approach would be more accurate, it leads to more complications and stiff source terms in the final evolution equations, which is why it is rarely used in the literature; see \cite{Palenzuela:2008sf,Dionysopoulou:2012zv,Wright:2019blb} for some examples. \\

In GRMHD, the full electromagnetic field is described by the Faraday tensor:
\begin{equation}
    F^{\mu\nu} = n^\mu E^\nu - n^\nu E^\mu + \epsilon^{\mu\nu\kappa\lambda} n_\kappa B_\lambda,
\end{equation}
where $\epsilon^{\mu\nu\kappa\lambda}$ is the four-dimensional Levi-Civita symbol and where $E^\nu$ and $B^\nu$ are, respectively, the electric and magnetic field as measured by an Eulerian observer. Accordingly, we can obtain the electric and magnetic field by the Faraday tensor and its dual $\sideset{^*}{}{\mathop F}^{\mu\nu} = \frac{1}{2} \epsilon^{\mu\nu\kappa\lambda} F_{\kappa\lambda}$ via:
\begin{equation}
    E^{\mu} = F^{\mu\nu} n_{\nu} \hspace{1cm} B^{\mu} = {}^{*}F^{\mu\nu} n_{\nu}.
\end{equation}

In the case of ideal GRMHD with infinite conductivity, there is no charge separation and the electric field vanishes in the fluids rest-frame. Thus, ideal GRMHD corresponds to imposing $u_\mu F^{\mu\nu}=0$ where $u_\mu$ is the fluid four-velocity. Following this condition, the electric field can be expressed by:
\begin{equation}
    E^{\nu} = -\frac{1}{W} \epsilon^{\mu\nu\kappa\lambda} n_\kappa B_\lambda u_\mu,
\end{equation}
where $W = \alpha u^0 = 1/\sqrt{1-v^2}$ is the Lorentz factor. It is therefore sufficient to evolve the magnetic field only, which greatly simplifies the final evolution equations. \par

We introduce the magnetic four-vector $b^\mu = u_\mu \sideset{^*}{}{\mathop F}^{\mu\nu}$ describing the magnetic field for a comoving observer. The components are given by:
\begin{equation}
    b^0 = \frac{W B^i v_i}{\alpha}, \hspace{1cm} b^i = \frac{B^i}{W} + b^0 (\alpha v^i - \beta^i),
\end{equation}
with $v^i = u^i / W + \beta^i / \alpha$ as the spatial components of the fluid velocity measured by an Eulerian observer. \par 

The full stress-energy tensor for a perfect fluid is then defined by:
\begin{equation}
    T^{\mu \nu} = \left( \rho h + b^2 \right) u^\mu u^\nu + \left(p + \frac{b^2}{2} \right) g^{\mu \nu} - b^\mu b^\nu,
\end{equation}
with $h = 1 + \epsilon + p/\rho$ being the specific enthalpy, $\rho$ the rest-mass density, $p$ the pressure, and $\epsilon$ the internal energy density. $b^2 = b^\mu b_\mu$ represents twice the magnetic pressure with $p_{\rm mag} = b^2/2$. We note that in our definition of the magnetic field a factor of $4 \pi$ is absorbed. \\

The evolution equations for ideal GRMHD are derived from the conservation laws of baryon number and energy-momentum:
\begin{equation}
    \nabla_\mu \left(\rho u^\mu\right) =0, \hspace{0.5cm} \nabla_\mu T^{\mu\nu} = 0,
    \label{eq:conslaw}
\end{equation}
as well as from the Maxwell's equations:
\begin{equation}
    \nabla_{\mu} {}^{*}F^{\mu \nu} = 0,
    \label{eq:Maxwell}
\end{equation}
with $\nabla_\mu$ being the covariant derivative. Furthermore, an EOS is required to close the evolution system. \par 
   
In order to write the evolution system in the form of a balance law with:
\begin{equation}
    \frac{\partial \textbf{q}}{\partial t} + \frac{\partial \textbf{F}^i}{\partial x^i} = \textbf{s},
    \label{eq:balancelaw}
\end{equation}
we define the primitive variables $\textbf{w}=\left(\rho, v^i, \epsilon, p, B^i \right)$ and the following conservative variables $\textbf{q}$:
\begin{align}
    D    =& \sqrt{\gamma} \rho W, \label{eq:D} \\
    S_j  =& \sqrt{\gamma}  \left(\left(\rho h +b^2\right) W^2 v_j- \alpha b^0 b_j\right), \label{eq:S} \\
    \tau =& \sqrt{\gamma}  \left(\left(\rho h +b^2\right) W^2 - \left(p+p_{\rm mag}\right)\right. \nonumber \\
           & \left.- \alpha^2 \left(b^0\right)^2 \right)- D, \label{eq:tau} \\
    \hat{B}^k  =& \sqrt{\gamma} B^k , \label{eq:B}
\end{align}     
where $\gamma$ is the determinant of $\gamma_{ij}$.
Finally, we obtain the following evolution equations in the form of Eq.~\eqref{eq:balancelaw}:
\begin{align}
   \textbf{q}   &= [D, S_j, \tau, \hat{B}^k], \\ 
   \textbf{F}^i &= \alpha \begin{bmatrix} 
                       D \tilde{v}^i \\ 
                       S_j \tilde{v}^i + \sqrt{\gamma} \left(p+p_{\rm mag}\right) \delta_j^i - b_j \hat{B}^i/W \\ 
                       \tau \tilde{v}^i + \sqrt{\gamma} \left(p+p_{\rm mag}\right) v^i - \alpha b^0 \hat{B}^i/ W \\
                       \hat{B}^k \tilde{v}^i - \hat{B}^i \tilde{v}^k \end{bmatrix}
   \label{fluxterm}, \\               
   \textbf{s} &= \alpha \sqrt{\gamma} \begin{bmatrix}  
                    0 \\
                    T^{\mu \nu} \displaystyle \left( \frac{\partial g_{\nu j}}{\partial x^\mu} - \Gamma^\lambda_{~\mu \nu} g_{\lambda j} \right)\\ 
                    \alpha \left(T^{\mu 0} \displaystyle \frac{\partial \ln (\alpha)}{\partial x^\mu}-T^{\mu \nu} \Gamma_{~\mu \nu}^0\right)\\ 
                    0^k                       
                    \end{bmatrix},
\end{align} 
with $\tilde{v}^i = v^i -\beta^i/\alpha$ and the Christoffel symbols $\Gamma_{~\mu \nu}^\lambda$.

\subsection{Divergence-Free Constraint Treatment}
\label{subsec:DivFree}

While the system of equations derived above is generally sufficient to perform simulations and evolve the magnetic field in time, there is no guarantee that the time component of Maxwell's equations \eqref{eq:Maxwell} is satisfied. Even when starting from constraint-satisfying initial data, numerical errors can accumulate leading to magnetic monopoles and thus to violations of Eq.~\eqref{eq:Maxwell}. Approaches typically used to address this issue include \textit{constraint transport} schemes as implemented in \gramx\ and \sacra, \textit{vector potential} methods as implemented in \spritz, and \textit{divergence cleaning} techniques. We refer to \cite{Shankar:2022ful,Kiuchi:2022ubj} and \cite{Cipolletta:2019geh} for implementations of \textit{constraint transport} and \textit{vector potential} methods in \gramx, \sacra, and \spritz, respectively. \par

Our GRMHD implementation follows the hyperbolic \textit{divergence cleaning} scheme from \cite{Liebling:2010bn,Mosta:2013gwu}. In particular, a new field variable $\zeta$ is introduced to damp and advect divergences of the magnetic field. We use a modification of the Maxwell's equations:
\begin{equation}
    \nabla_\mu \left( {}^*F^{\mu \nu} + g^{\mu \nu} \zeta \right) = \kappa n^\nu \zeta,
    \label{eq:modMaxwell}
\end{equation}
with the damping rate $\kappa$. We set $\kappa = 1$ in this work. For $\zeta \rightarrow 0$, the equation reduces again to the Maxwell's equations \eqref{eq:Maxwell}. From the time component of Eq.~\eqref{eq:modMaxwell}, we obtain an evolution equation for $\zeta$:
\begin{align}
    \partial_t \zeta &+ \partial_i \left( \frac{\alpha}{\sqrt{\gamma}} \hat{B}^i - \zeta \beta^i \right) \nonumber\\
    &= \zeta (-\kappa \alpha - \partial_i \beta^i) + \hat{B}^i \partial_i \left(\frac{\alpha}{\sqrt{\gamma}}\right).
\end{align}

The spatial part of Eq.~\eqref{eq:modMaxwell} leads to the following modification for the evolution equation of the magnetic field:
\begin{align}
   \partial_t \hat{B}^j &+ \partial_i \left[ \left(\alpha v^i - \beta^i \right) \hat{B}^j - \alpha v^j \hat{B}^i + \alpha \sqrt{\gamma} \gamma^{ij} \zeta \right] \nonumber \\
   &= - \hat{B}^i \partial_i \beta^j + \zeta \partial_i \left( \alpha \sqrt{\gamma} \gamma^{ij}\right).
\end{align}

Although divergences of the magnetic field are damped, this approach does not prevent the occurrence of constraint violations, unlike the \textit{constraint transport} or \textit{vector potential} methods. This is an obvious disadvantage of the method, leading to finite errors. Nonetheless, these errors should converge to zero in a reliable and controlled manner with the scheme employed, similar to the convergence of other constraints, e.g., the Hamiltonian constraint. On the other hand, the implementation is much simpler as there is no staggering of the magnetic field necessary and available high-order HRSC schemes can be directly employed. While \bam\ already uses cell-centered fields for both the metric and fluid variables, additional work would be required since the magnetic field is defined at the cell faces and not in the cell centers. In particular, the AMR restriction and prolongation algorithms would have to be changed and would become more complex. For a magnetic field defined at the cell surface, it would first have to be interpolated to the cell centers. In most codes using \textit{constraint transport} and \textit{vector potential} methods, this is often done linearly and, thus, the magnetic field relies on second-order methods. For a cell-centered magnetic field, however, we can use the same methods as for other variables, typically using fourth-order schemes in our implementation. Apart from questions of complexity and convergence rate, a key question is how a given physical system depends on numerical errors in the divergence constraint, which we investigate in the numerical examples that follow.

\subsection{Numerical Schemes}

\subsubsection{Conversion to Primitive Variables}

In order to write the evolution equations in form of a balance law, we use the conservative variables $\textbf{q}$ as defined in Eqs.~(\ref{eq:D})--(\ref{eq:B}) which are analytical functions of the primitive variables $\textbf{w}$. Nevertheless, we need to know the primitive variables, for instance, to compute the flux terms $\textbf{F}^i$ at each time step. However, the inversion of Eqs.~(\ref{eq:D})--(\ref{eq:B}) to obtain the primitive variables is non-trivial and must be solved numerically. We refer to \cite{Siegel:2017sav} for a summary and discussion of different approaches commonly used in NR codes. The scheme we use follows the approach implemented in the \textsc{RePrimAnd} library~\cite{Kastaun:2020uxr}. More precisely, we use its master-function (see Eq.~(44) of Ref.~\cite{Kastaun:2020uxr}) and use a Brent-Dekker scheme to find its root. In the numerical tests as well as in the BNS merger simulations, which we discuss in Secs.~\ref{sec:FirstTests} and \ref{sec:BNSSimulations}, this method has proven to be robust and provides reliable results.

\subsubsection{Reconstruction}
\label{subsubsec:reconstruction}

\bam\ contains HRSC methods for the treatment of shocks and discontinuities in the hydrodynamic variables. In order to evaluate the fluxes $\textbf{F}$ at the individual grid cell interfaces and to solve the Riemann problems, the variables have to be reconstructed to the cell surfaces from either side. Since all field variables are cell-centered in our GRMHD implementation, we can use the same reconstruction schemes for the magnetic field variables and $\zeta$ that are already implemented in \bam. These include, for example, a third-order convex-essentially-non-oscillatory (CENO3) scheme~\cite{Liu:1998,Zanna:2002qr}, a fifth-order weighted-essentially-non-oscillatory (WENOZ) scheme~\cite{Borges:2008}, or a fifth-order monotonicity preserving (MP5) scheme ~\cite{Suresh:1997}. \par 

While higher order schemes are generally more accurate and less dissipative, they can exhibit stronger oscillations, as we show and discuss in some relativistic shock tests in Sec.~\ref{sec:FirstTests}. Enhanced Oscillations can lead to unphysical values. For this reason, in the BNS merger simulations with \bam\ presented in Sec.~\ref{sec:BNSSimulations}, we apply the following procedure in lower density regimes to ensure the physical validity of reconstructed variables and, in particular, to maintain the positivity of the density and pressure:

\begin{itemize}
    \item We use a high-order reconstruction scheme to compute the GRMHD variables at the interface between grid cell $i$ and $i+1$. 
    \item If the rest-mass density $\rho$ drops below a threshold density, we examine the oscillation of all variables by comparing the reconstructed values at the interface with the cell-centered ones of $i$ and $i+1$. 
    \item Whenever one reconstructed variable is below or above a value in the grid cells $i$ and $i+1$, we change to a low-order reconstruction and recompute the values of all variables at the interface.
    \item Finally, we examine the physical validity of the reconstructed values by demanding a positive rest-mass density and a positive pressure. If this is not given, we use a linear reconstruction method.
\end{itemize}

For high-order reconstruction schemes, we typically use WENOZ or MP5 as fifth-order methods or CENO3 as third-order method. We then fall back on a linear total variation diminishing (TVD) method with ``minmod'' slope limiter \cite{Toro} as low-order reconstruction scheme.

\subsubsection{Riemann Solver}
\label{subsubsec:riemannsolver}

For the GRMHD simulations with \bam\ in Secs.~\ref{sec:FirstTests} and \ref{sec:BNSSimulations}, we use the Harten, Lax, and van Leer (HLL) Riemann solver \cite{Harten:1983}. HLL uses a two-wave approximation via estimates for the fastest left- and right-moving signal speeds:
\begin{align}
    \textbf{F}^{\rm HLL} &= \frac{\lambda_{\rm max}\textbf{F}\left(\textbf{q}^{+}\right) - \lambda_{\rm min}\textbf{F}\left(\textbf{q}^{-}\right)}{\lambda_{\rm max} - \lambda_{\rm min}} \nonumber \\
    &- \frac{\lambda_{\rm max}\lambda_{\rm min}}{\lambda_{\rm max} - \lambda_{\rm min}}\left(\textbf{q}^{+}-\textbf{q}^{-}\right),
\end{align}
where $\textbf{q}^{+}$ and $\textbf{q}^{-}$ are the variables respectively reconstructed from the left and right side of the cell face, and $\textbf{F}\left(\textbf{q}^{+}\right)$ and $\textbf{F}\left(\textbf{q}^{-}\right)$ are the according flux terms. The fastest left- and right-moving signal speeds are determined by:
\begin{align}
    \lambda_{\rm max} &= \max \left(\lambda_i\left(\textbf{q}^{+}\right),\lambda_i\left(\textbf{q}^{-}\right),0\right), \\
    \lambda_{\rm min} &= \min \left(\lambda_i\left(\textbf{q}^{+}\right),\lambda_i\left(\textbf{q}^{-}\right),0\right), 
\end{align}
with $\lambda_i$ being the characteristic wave speeds. \par

While there are three characteristic modes in pure general relativistic hydrodynamics, there are seven independent characteristic waves in ideal GRMHD: the entropy wave, two Alfven waves, and four magnetosonic waves (two slow and two fast modes). For a detailed discussion of the individual waves and derivation of the eigenvalues, we refer to \cite{Anton:2005gi}. Since the exact wave speeds require the solution of a non-trivial quartic equation for the magnetosonic waves, we use the commonly applied approximation proposed in \cite{Gammie:2003rj}. We note that by using \textit{divergence cleaning}, two additional modes with characteristic velocities equal to the speed of light must be taken into account. For the magnetic field variables and $\zeta$, we therefore set $\lambda_{\rm max}=1$ and $\lambda_{\rm min}=-1$ in the Riemann solver. \\

In addition to HLL, we started to implement the HLLD, a five-wave relativistic Riemann solver, following \cite{Mignone:2008ii,Kiuchi:2022ubj}, with minor changes to adapt to the \textit{divergence cleaning} formulation. The implementation is still in test phase. For this reason, we use it only in some of the special relativistic tests presented in Sec.~\ref{sec:FirstTests}.

\subsubsection{Atmosphere}
\label{subsubsec:atmosphere}

In grid-based NR simulations, the vacuum region surrounding the compact objects is usually modeled with an artificial atmosphere. The reason is that too low rest-mass densities are numerically very demanding and can lead to errors in the recovery of the primitive variables and in HRSC schemes. Therefore, we set in \bam\ a grid cell to atmosphere when its density falls below a density threshold $\rho_{\rm thr}$ \cite{Thierfelder:2011yi}. We assume a cold, static atmosphere: the density is reduced to a fraction of the star's original central density $\rho_{\rm atm} = f_{\rm atm} \rho_c$, pressure and internal energy are determined according to the zero temperature part of the EOS, and the velocity of the fluid is set to zero. The magnetic field remains unchanged. The density threshold is defined as the fraction of the atmosphere value with $\rho_{\rm thr} = f_{\rm thr} \rho_{\rm atm}$ to avoid fluctuations around the density floor. In the BNS merger simulations presented in Sec.~\ref{sec:BNSSimulations}, we use $f_{\rm atm} = 10^{-11}$ and $f_{\rm thr} = 100$.

\section{Special Relativistic Tests}
\label{sec:FirstTests}

For validating our new implementation, we have performed a series of well-established special relativistic magneto-hydrodynamics tests in one-, two-, and three-dimensions. In particular, it is important to analyze the reliability of the code with respect to shocks. In each test, we compare our results and performance with the codes \spritz~\cite{Cipolletta:2019geh}, \sacra~\cite{Kiuchi:2022ubj}, and \gramx~\cite{Shankar:2022ful}.

\subsection{One-Dimensional Tests}

\begin{figure*}[t]
    \centering
    \includegraphics[width=\linewidth]{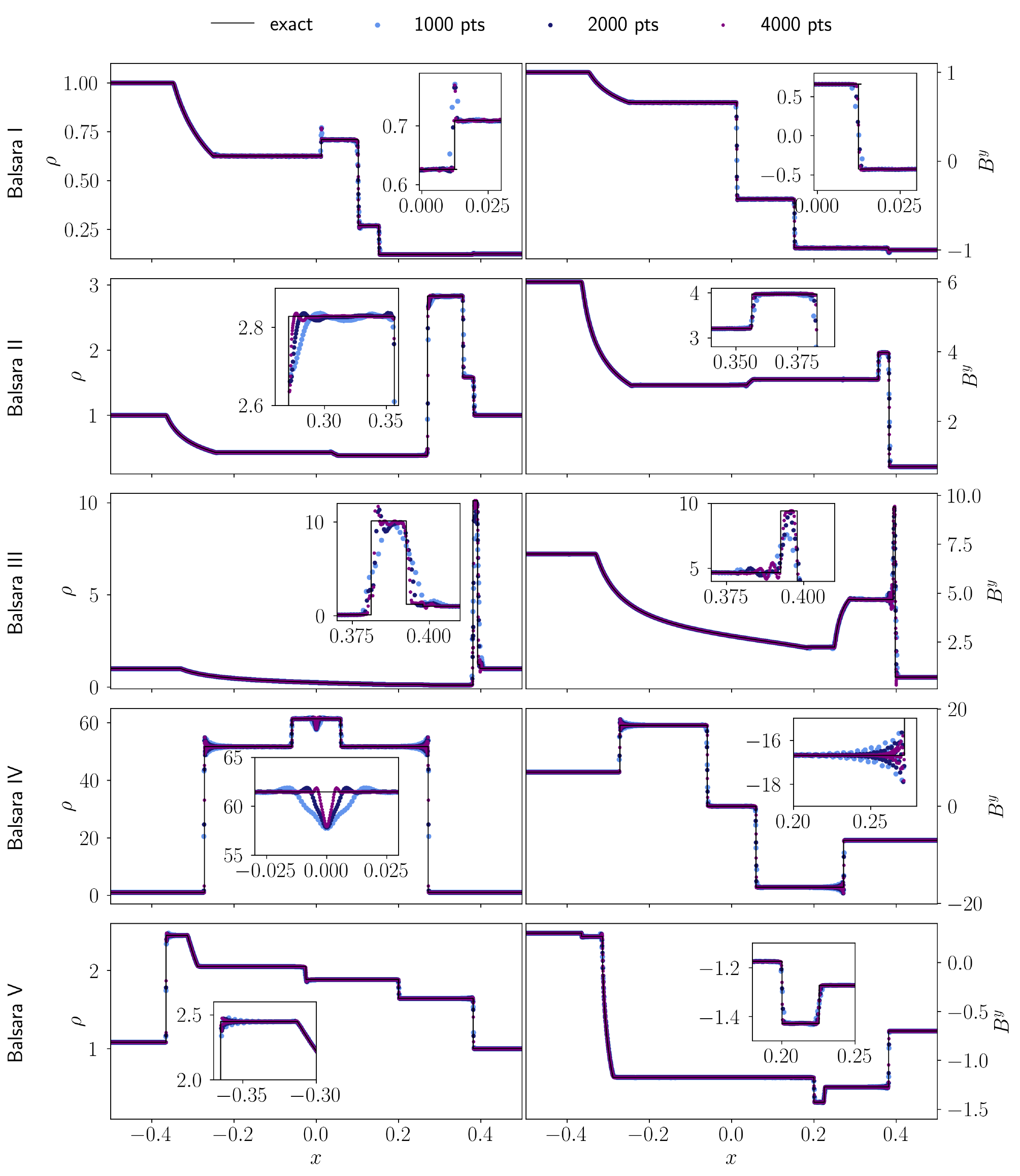}
    \caption{Results of Balsara tests I to V (from top to bottom) for \bam. We compare the numerical results for three different resolutions using $1000$, $2000$, and $4000$~grid points with the exact solution of \cite{Giacomazzo:2005jy} (continuous black lines). Left and right columns show respectively the profiles of density $\rho$ and magnetic field component $B^y$ at $t=0.4$ for Balsara tests I--IV and at $t=0.55$ for Balsara test V. All tests are performed using the MP5 reconstruction method.}
    \label{fig:balsara}
\end{figure*}

\begin{figure}[t]
    \centering
    \includegraphics[width=\linewidth]{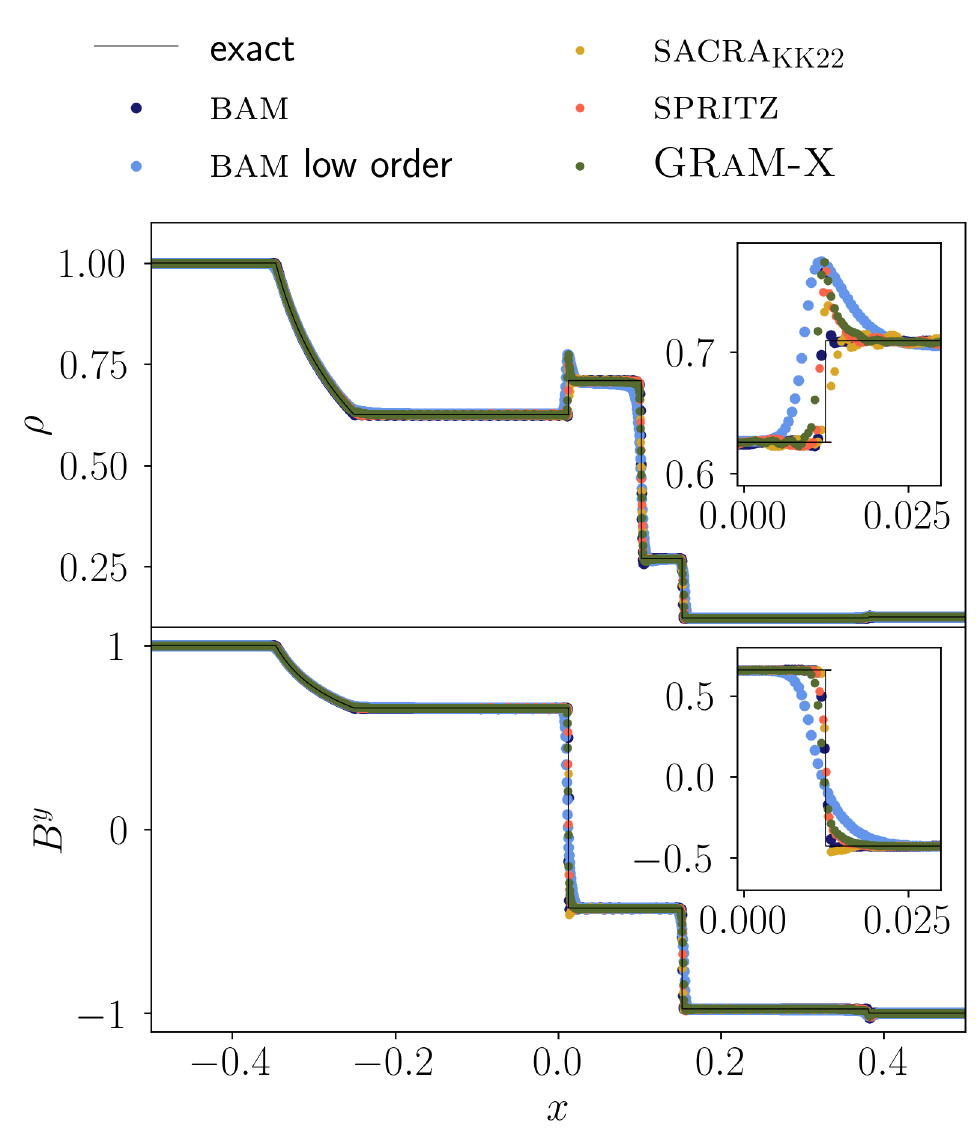}
    \caption{Comparison of Balsara test I performed with \bam, \sacra, \spritz, and \gramx. We show the profiles of $\rho$ and $B^y$ at $t=0.4$ for one resolution using $2000$~grid points and add the exact solution as continuous black lines. We note that the codes use different reconstruction schemes: \gramx\  a fifth-order WENO5, \spritz\ a third-order PPM, and \sacra\ a third-order PPM scheme. For \bam,  we show the results once using MP5 and once using a linear TVD method (labeled as `low order').}
    \label{fig:balsara_I}
\end{figure}

\begin{figure}[t]
    \centering
    \includegraphics[width=0.97\linewidth]{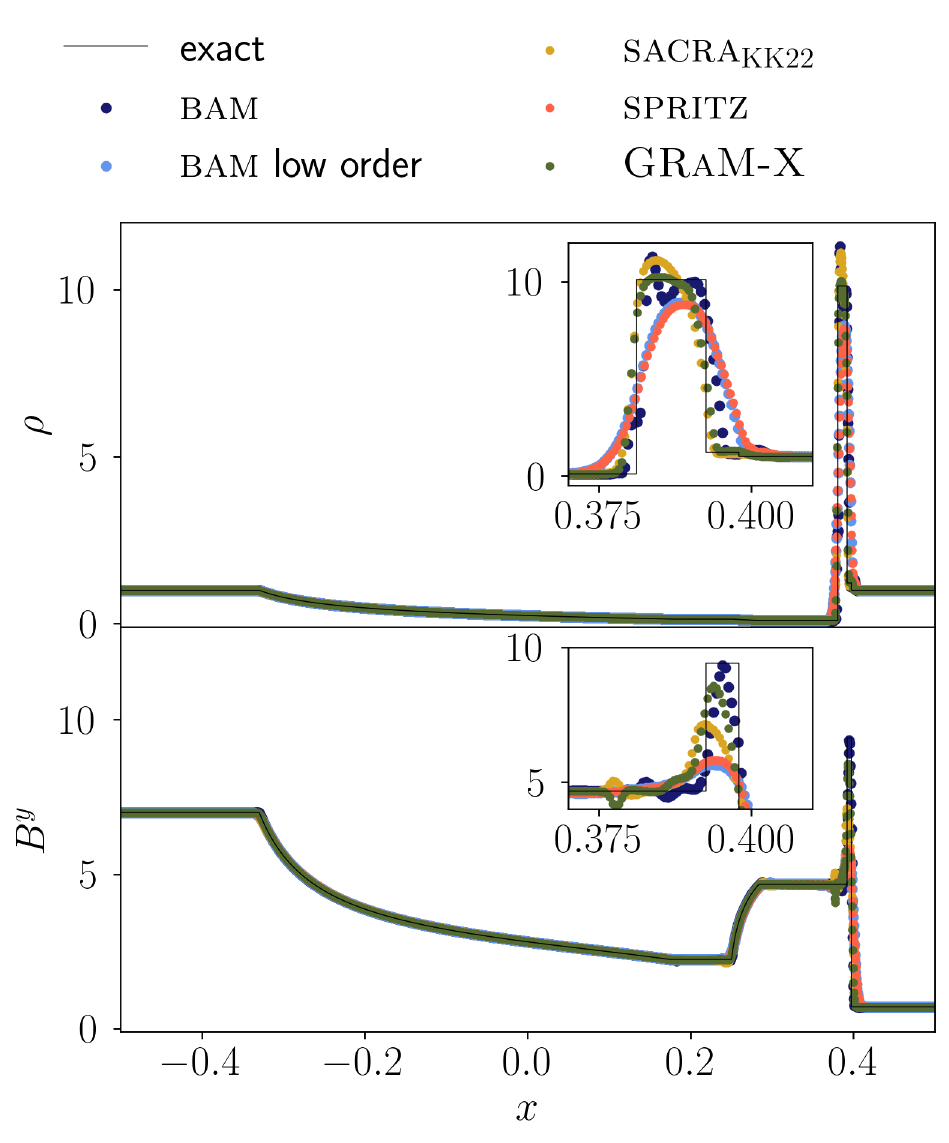}
    \caption{Comparison of Balsara test III performed with \bam, \sacra, \spritz, and \gramx. We show the profiles of $\rho$ and $B^y$ at $t=0.4$ for one resolution using $2000$~grid points and add the exact solution as continuous black lines. We note that the codes use different reconstruction schemes: \gramx\ a fifth-order WENO5, \spritz\ a second-order TVD, and \sacra\ a third-order PPM scheme. For \bam, we show the results once using MP5 and once using a linear TVD method (labeled as `low order').}
    \label{fig:balsara_III}
\end{figure}

Relativistic shock tube problems are simple, one-dimensional tests that can be used to demonstrate a code's ability to capture a variety of different shock wave structures. By setting different states of the fluid to the left and right of an interface, a shock is generated and evolved. Among the most popular and well-known shock tube problems in magneto-hydrodynamics are the Balsara tests~\cite{Balsara:2001}. We summarize the initial data for the Balsara tests I--V in Tab.~\ref{tab:Balsara} by specifying the respective fluid variables of the left and right states. The tests are performed along the $x$-axis with the initial fluid discontinuity at $x=0$. In all Balsara tests, an ideal-gas law EOS with $p=(\Gamma-1)\rho\epsilon$, where $\Gamma$ is the adiabatic constant, is assumed: in Balsara test I with $\Gamma = 2.0$ and in Balsara test II--V with $\Gamma = 5/3$. We evolve the shock waves until $t=0.4$ for Balsara tests I--IV and $t=0.55$ for Balsara test V. \par 

The results are shown in Fig.~\ref{fig:balsara} for each Balsara test at three different resolutions with $1000$, $2000$, and $4000$~grid points in $x \in \left[-0.5, 0.5\right]$. We use MP5 reconstruction. For each test, we show the profiles of the density $\rho$ and the $B^y$ component of the magnetic field at the respective final time. Additionally, we show the exact solution of \cite{Giacomazzo:2005jy} as comparison. Overall, all tests are in good agreement. As expected, higher resolutions increase the accuracy and are better able to capture the shock fronts. This is particularly evident in the Balsara tests II, III and IV, where a high resolution is required to capture the sharp edges of the shock waves in $\rho$. Some oscillations are visible, e.g., in $B^y$ of Balsara tests III and IV, that we attribute to the use of a reconstruction scheme with relatively high order. \par

In order to evaluate how our GRMHD implementation performs in these tests with respect to other NR codes, we compare the results obtained with \bam\ with those obtained with \sacra, \gramx, and \spritz. Figures~\ref{fig:balsara_I} and \ref{fig:balsara_III} show respectively the Balsara tests I and III for all four codes. The resolution is set to $2000$~grid points in $x \in \left[-0.5, 0.5\right]$. Although we try to choose the technical configurations rather similar, each working group using different codes has preferred numerical methods that also differ in the individual implementations. For instance, the HLLD Riemann solver as described in \cite{Kiuchi:2022ubj} is used for both tests by \sacra. This solver is more advanced than the HLL solver used by \bam, \spritz, and \gramx. Furthermore, different methods are used to reconstruct the fluid variables: While \gramx\ uses a fifth-order WENO5 scheme, \sacra\ applies a third-order piecewise parabolic method (PPM). \spritz\ also employs PPM reconstruction in Balsara test I. However, since Balsara test III is more demanding due to the large jump in the initial pressure, \spritz\ uses here a second-order TVD method with ``minmod'' limiter. To assess the differences caused by different reconstruction algorithms, we show for \bam\ results obtained with a linear TVD method, labeled as `low order' in Figs.~\ref{fig:balsara_I} and \ref{fig:balsara_III}, additionally to those with MP5 reconstruction. \par 

All codes demonstrate their ability to reliably capture and evolve the shock waves with small differences at the shock fronts. For Balsara test III, \gramx, \sacra, and \bam\ (with MP5) seem to be superior in modeling the exact solution, despite showing small oscillations. We explain this by the usage of higher orders for the reconstruction. \spritz\ uses a second-order scheme in this test. In fact, the solution of \bam\ with linear TVD method perfectly matches the results of \spritz\ when the same reconstruction method and Riemann solver are used. Also in Balsara test I, the largest differences appear for \bam\ using different reconstruction methods. This suggest that the differences in Figs.~\ref{fig:balsara_I} and \ref{fig:balsara_III} for the individual codes are due to different HRSC algorithms rather than different methods for the magnetic field or the divergence-free constraint treatments.

\begin{table}[t]
    \centering
    \caption{Initial data for Balsara tests I-V. The shock tube tests are performed along the $x$-axis. We list for each test the left and right states of the fluid: the rest-mass density $\rho$, the pressure $p$, the velocity components $v^x$, $v^y$, and $v^z$, and the magnetic field components $B^x$, $B^y$, and $B^z$.}
    \label{tab:Balsara}
    \begin{tabular}{cc|cccccccc}
    \hline
     \multicolumn{2}{c}{Balsara} & $\rho$ & $p$ & $v^x$ & $v^y$ & $v^z$ & $B^x$ & $B^y$ & $B^z$ \\ 
     \hline \hline
     I   & left  & $1.0$   & $1.0$    & $0.0$    & $0.0$  & $0.0$ & $0.5$  & $1.0$  & $0.0$ \\
         & right & $0.125$ & $0.1$    & $0.0$    & $0.0$  & $0.0$ & $0.5$  & $-1.0$ & $0.0$ \\
     \hline
     II  & left  & $1.0$   & $30.0$   & $0.0$    & $0.0$  & $0.0$ & $5.0$  & $6.0$  & $6.0$ \\
         & right & $1.0$   & $1.0$    & $0.0$    & $0.0$  & $0.0$ & $5.0$  & $0.7$  & $0.7$ \\
     \hline
     III & left  & $1.0$   & $1000.0$ & $0.0$    & $0.0$  & $0.0$ & $10.0$ & $7.0$  & $7.0$ \\
         & right & $1.0$   & $0.1$    & $0.0$    & $0.0$  & $0.0$ & $10.0$ & $0.7$  & $0.7$ \\
     \hline
     IV  & left  & $1.0$   & $0.1$    & $0.999$  & $0.0$  & $0.0$ & $10.0$ & $7.0$  & $7.0$ \\
         & right & $1.0$   & $0.1$    & $-0.999$ & $0.0$  & $0.0$ & $10.0$ & $-7.0$ & $-7.0$ \\
     \hline
     V   & left  & $1.08$  & $0.95$   & $0.4$    & $0.3$  & $0.2$ & $2.0$  & $0.3$  & $0.3$ \\
         & right & $1.0$   & $1.0$    & $-0.45$  & $-0.2$ & $0.2$ & $2.0$  & $-0.7$ & $0.5$ \\
     \hline
    \end{tabular}
\end{table}

\subsection{Two-Dimensional Tests}

\subsubsection{Cylindrical Explosion}
\label{subsubsec:CE}

\begin{figure*}[t]
    \centering
    \includegraphics[width=\linewidth]{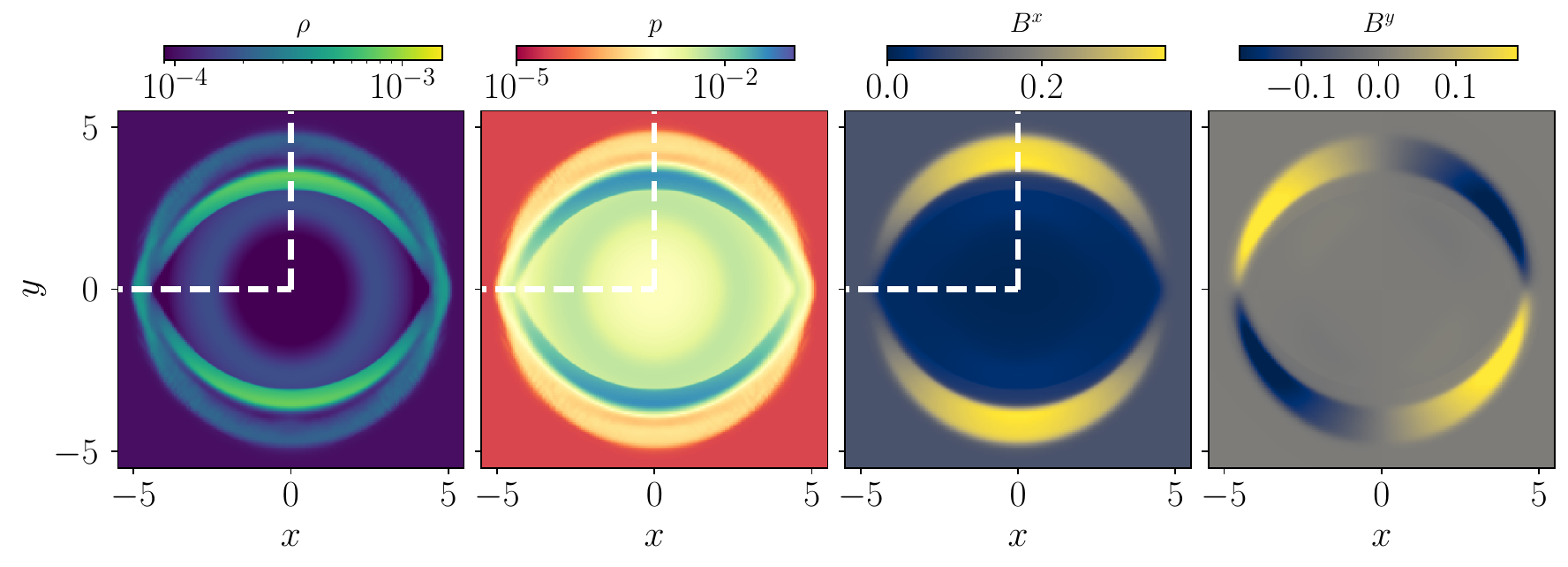}
    \caption{Snapshots of $\rho$, $p$, $B^x$, and $B^y$ for the cylindrical explosion test with \bam\ at the final time $t=4$. The white dashed lines mark the one-dimensional profiles shown in Fig.~\ref{fig:CE1D_nomesh}.}
    \label{fig:cylindricalexplosion}
\end{figure*}

\begin{figure}[t]
    \centering
    \includegraphics[width=0.9\linewidth]{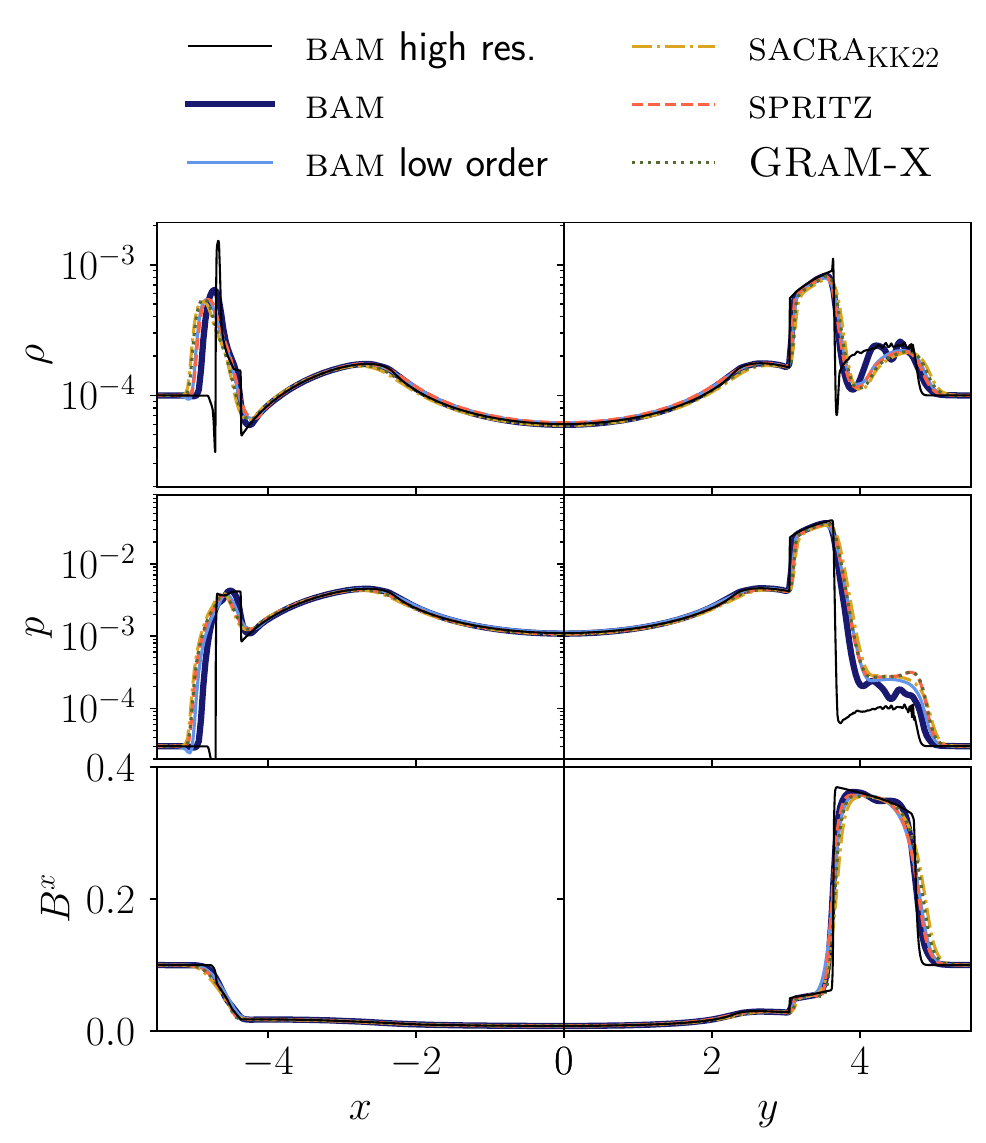}
    \caption{One-dimensional cuts along negative $x$-axis and positive $y$-axis of the cylindrical explosion test with \bam, \spritz, \gramx, and \sacra. We show for \bam\ results with CENO3/linear TVD  and with linear TVD (labeled as `low order') reconstruction as well as high-resolution results using one refinement level with $8000 \times 8000$~grid points (labeled as `high res.'). In \gramx\ and \spritz\ a second-order TVD, and in \sacra\ a third-order PPM reconstruction are used.}
    \label{fig:CE1D_nomesh}
\end{figure}

The first two-dimensional shock problem we perform in this test series is the cylindrical explosion, also known as cylindrical blast wave, a standard problem for GRMHD codes, e.g., \cite{Komissarov:1999,DelZanna:2007pk,Etienne:2010ui,Mosta:2013gwu}. We consider a dense medium within a cylindrical radius of $r_{\rm in} = 0.8$ in a less dense ambient medium outside of $r_{\rm out} =1$. The denser medium has an overpressure and therefore expands. The initial values for density $\rho$ and pressure $p$ are: $\rho_{\rm in } = 10^{-2}$ and $p_{\rm in } = 1.0$  for the inner medium, and $\rho_{\rm out } = 10^{-4}$ and $p_{\rm out } = 3 \times 10^{-5}$ for the outer medium. For the transition region $0.8 \leq r \leq 1$, we apply:
\begin{align}
    \rho &= \exp \left( \frac{ \left(r_{\rm out} -r \right) \ln \rho_{\rm in} + \left(r - r_{\rm in} \right) \ln \rho_{\rm out} }{r_{\rm out}-r_{\rm in}} \right), \\
    p &= \exp \left( \frac{ \left(r_{\rm out} -r \right) \ln p_{\rm in} + \left(r - r_{\rm in} \right) \ln p_{\rm out} }{r_{\rm out}-r_{\rm in}} \right),
\end{align}
with cylindrical radius $r = \sqrt{x^2 + y^2}$ in the $x$-$y$ plane. In this test, an ideal-gas EOS with adiabatic index $\Gamma = 4/3$ is used and we set a uniformly directed magnetic field along the $x$-axis with $B^x = 0.1$ and $B^y = B^z = 0$. We evolve this configuration until $t=4$. \par 

The resulting blast wave at the final time is presented in Fig.~\ref{fig:cylindricalexplosion}. We show snapshots of $\rho$, $p$, $B^x$, and $B^y$ in the $x$-$y$ plane of a simulation with \bam\ using CENO3 reconstruction and a grid consisting of two refinement levels with $200 \times 200$~grid points each, covering a range of $\left[-6, 6\right]$ in $x$ and $y$ directions. A fast forward shock and a reverse shock bounding the inner region are observable. While there is no exact solution available for this tests, the structures shown in our results are consistent with those in \cite{Komissarov:1999,Mosta:2013gwu,Cipolletta:2019geh}. \par 

For a more quantitative comparison of our results with \spritz, \sacra, and \gramx, we show one-dimensional slices of the final blast wave for $\rho$, $p$, and $B^x$ along the $x$- and $y$-axes in Fig.~\ref{fig:CE1D_nomesh}. To avoid discrepancies due to different algorithms for flux conservation at the refinement boundaries, we use here a grid consisting of a single refinement level with grid spacing $\Delta x = 0.03$ and $400 \times 400$~points. For \bam, we apply the reconstruction as described in Sec.~\ref{subsubsec:reconstruction} with CENO3 as higher order scheme and linear TVD as lower order method, examining oscillations in the fluid variables in regions with $\rho < 2\times 10^{-4}$. Additionally, we run \bam\ once using a pure linear TVD method, labeled as `low order' in Fig.~\ref{fig:CE1D_nomesh}. Both, \gramx\ and \spritz, apply here a second-order TVD scheme, while \sacra\ results are shown for third-order PPM reconstruction. Lacking a well-known analytical solution for this test problem, we perform additionally one high-resolution test run with \bam\ using $8000 \times 8000$~grid points, i.e., a spacing of $\Delta x = 0.0015$, using CENO3. Except for \sacra, that applies the HLLD Riemann solver, the results are obtained with HLL. The results for \sacra, \gramx, and \spritz\ overlap with the ones obtained with \bam\ using the linear TVD reconstruction method. The higher-order reconstruction leads to stronger oscillations at the shock fronts, which are mainly visible in the $\rho$ and $p$ profiles along the $y$-axis. Nevertheless, all codes converge approximately to the same solution. As before in the Balsara tests, we conclude that the differences are primarily due to different HRSC schemes.

\subsubsection{Magnetic Rotor}

\begin{figure*}[t]
    \centering
    \includegraphics[width=\linewidth]{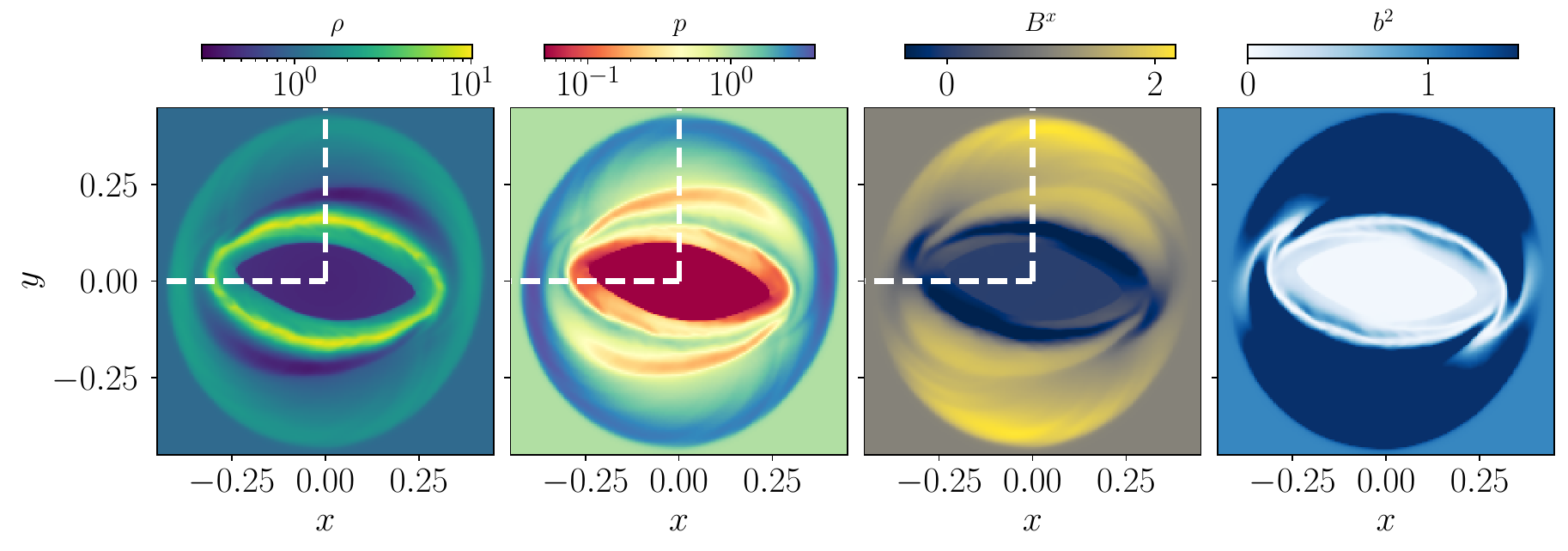}
    \caption{Snapshots of $\rho$, $p$, $B^x$, and $b^2$ for the magnetic rotor test with \bam\ at the final time $t=0.4$. The white dashed lines mark the one-dimensional profiles shown in Fig.~\ref{fig:MR1D_nomesh}.}
    \label{fig:magneticrotor}
\end{figure*}

\begin{figure}[t]
    \centering
    \includegraphics[width=0.9\linewidth]{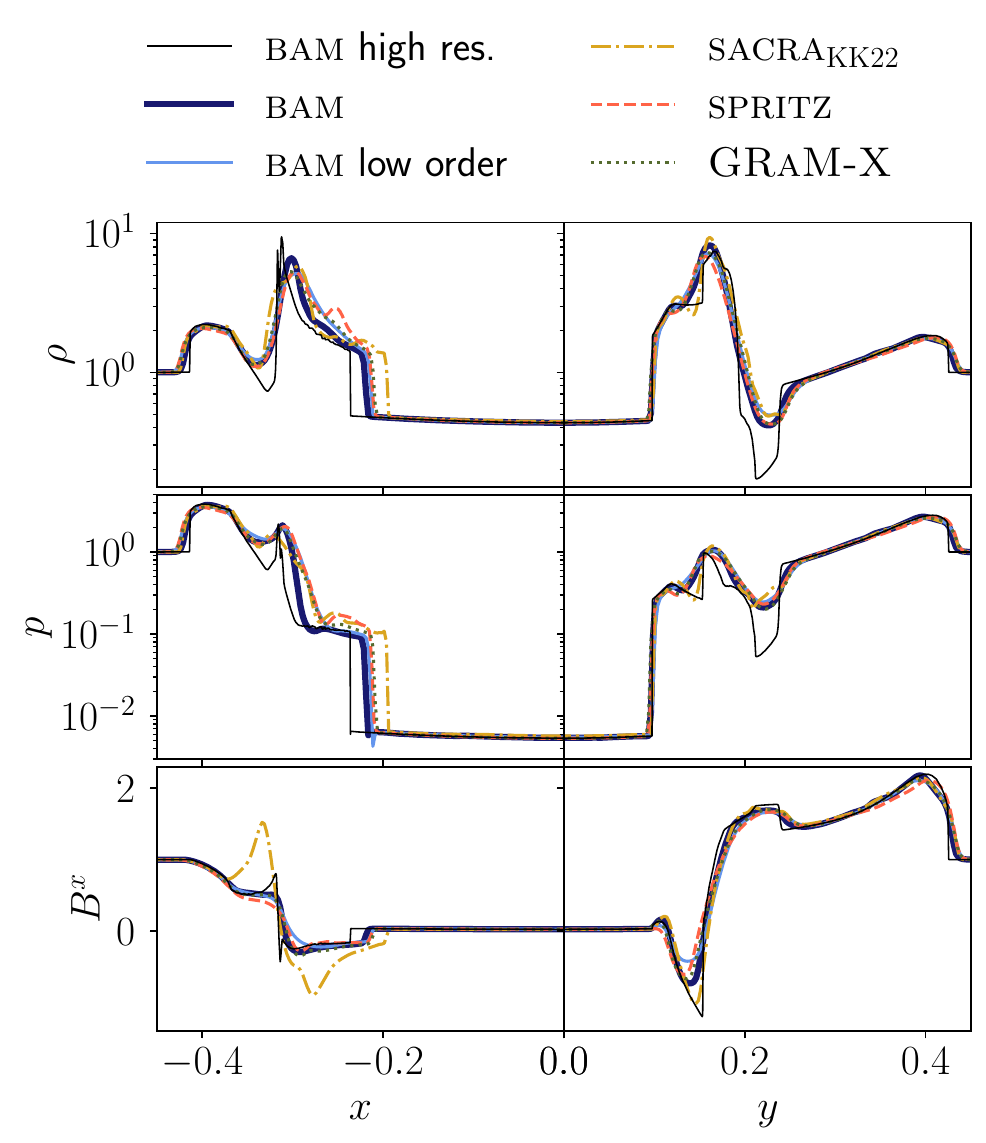}
    \caption{One-dimensional cuts along negative $x$ and positive $y$ axis of the magnetic rotor test with \bam, \spritz, \gramx, and \sacra. We show for \bam\ results with CENO3 and with linear TVD (labeled as `low order') reconstruction as well as high-resolution results using one refinement level with $8000 \times 8000$~grid points (labeled as `high res.'). In \gramx\ and \spritz\ a second-order TVD, and in \sacra\ a third-order PPM reconstruction are used.}
    \label{fig:MR1D_nomesh}
\end{figure}

As a second two-dimensional standard test, we consider the magnetic rotor, originally described for classical magneto-hydrodynamics \cite{Balsara:1999,Toth:2000} and later generalized to the relativistic case \cite{Zanna:2002qr, Etienne:2010ui}. For this test, we initialize a dense, fast rotating fluid in the center of a non-rotating ambient medium. The central rotating fluid is set inside an area with cylindrical radius $r=0.1$ with $\rho_{\rm in} = 10$ and a uniform angular velocity of $\Omega = 9.95$, resulting in a maximum velocity of the fluid of $v_{\rm max} = 0.995$. The density of the surrounding medium is $\rho_{\rm out} = 1$. We apply in this test an ideal-gas EOS with $\Gamma = 5/3$ and consider a uniform pressure with $p=1$ and magnetic field with $B^x = 1$, and $B^y=B^z=0$. The test is simulated until $t=0.4$. \par

We perform this test with \bam\ on a grid with two refinement levels and $200 \times 200$~points covering the range $\left[-0.5, 0.5 \right]$ for $x$ and $y$ using CENO3 reconstruction. During the simulation, the rotation leads to a twisting of the magnetic field lines in the central region, which in turn slows down the rotor. Figure~\ref{fig:magneticrotor} presents maps of the density $\rho$, the pressure $p$, and the magnetic field variables $b^2$, which can be considered as twice the magnetic pressure, and the $B^x$ component at the final time. The results are in good agreement with those in \cite{Zanna:2002qr,Etienne:2010ui,Mosta:2013gwu}. \par 

Like for the cylindrical explosion, we compare our results with \spritz, \sacra, and \gramx\ for one-dimensional slices at the final time. Figure~\ref{fig:MR1D_nomesh} shows profiles of $\rho$, $p$, and $B^x$ along the $x$- and $y$-axes. As for the cylindrical explosion, we choose for the comparison a grid without mesh refinement consisting of $400 \times 400$~grid points with a spacing of $\Delta x = 0.0025$. The results with \sacra\ are obtained with the HLLD Riemann solver in combination with a third-order PPM reconstruction, while \spritz\ and \gramx\ apply HLL and a second-order TVD reconstruction. For \bam\ we show the results using the HLL Riemann solver once with CENO3 and once with linear TVD, labeled as `low order' in Fig.~\ref{fig:MR1D_nomesh}. Again, we perform additionally one high-resolution test run for \bam\ with HLL and CENO3 using $8000 \times 8000$~grid points and a spacing of $\Delta x = 0.000125$. 

\begin{figure}[t]
    \centering
    \includegraphics[width=0.9\linewidth]{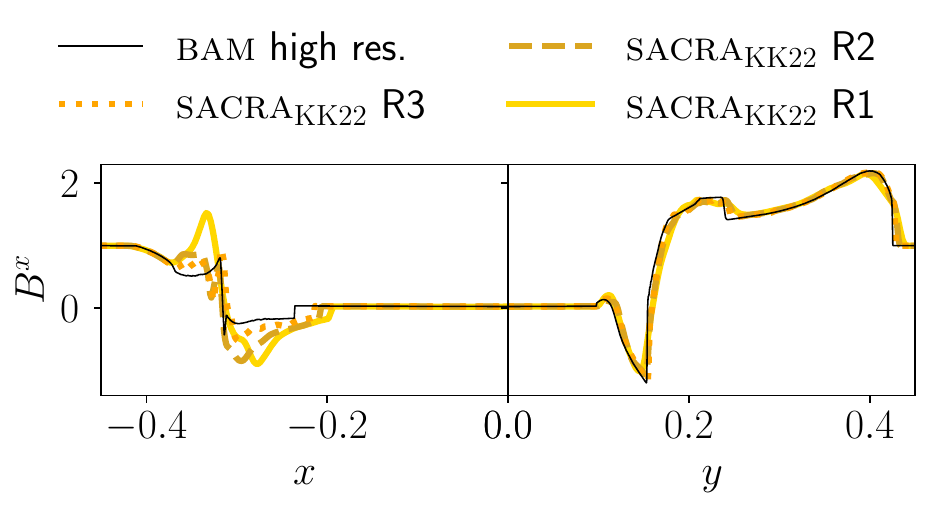}
    \caption{One-dimensional cuts for $B^x$ along negative $x$ and positive $y$ axis of the magnetic rotor test with \sacra\ for different resolutions: R1 with $\Delta x = 0.0025$, R2 with $\Delta x = 0.00125$, and R3 with $\Delta x = 0.000625$. We show additionally the high resolution results for \bam\ with $\Delta x = 0.000125$.}
    \label{fig:MR1D_SACRA}
\end{figure}

Overall, the results of the individual codes agree well. Mostly at the shock fronts, i.e., along $x$ between $-0.2$ and $-0.35$ and along $y$ between $0.1$ and $0.25$, some codes show stronger oscillations than others, which is strongly depending on the HRSC method used. The largest differences are present for \sacra\ for the $B^x$ profile. However, the differences decrease for higher resolutions. As shown in Fig.~\ref{fig:MR1D_SACRA}, with increasing resolution from $\Delta x = 0.0025$ to $\Delta x = 0.00125$ up to $\Delta x = 0.000625$ the $B^x$ profile for \sacra\ converges against the same solution. The shock along the $y$-axis is well captured even at the lowest resolution, while the shock along the $x$-axis needs a higher resolution. Since the main difference between \sacra\ and the other codes is the Riemann solver, we attribute to the HLLD solver being more suitable for different types of shock waves than for others.

\subsubsection{Kelvin Helmholz Instability}

\begin{figure}[t]
    \centering
    \includegraphics[width=\linewidth]{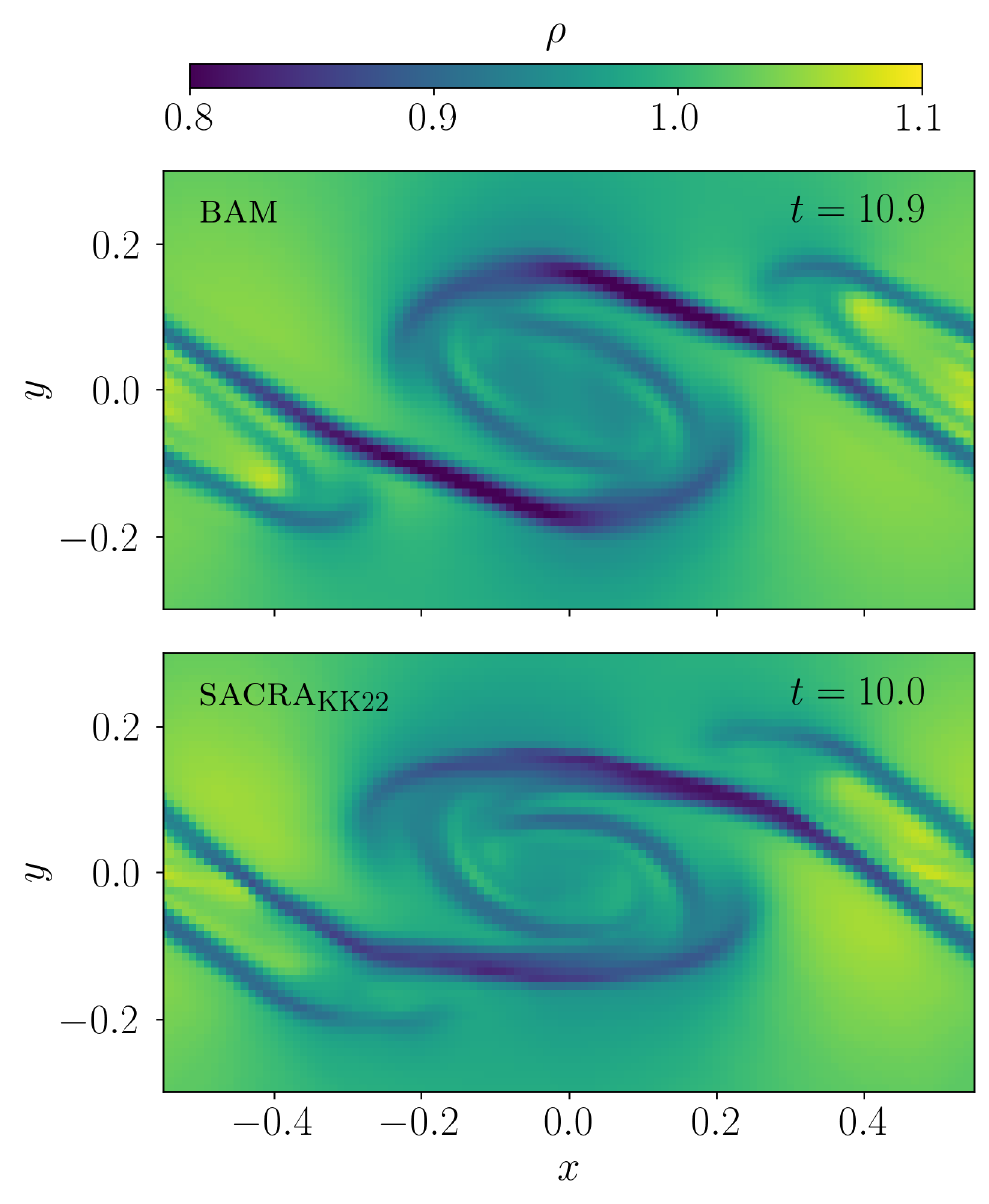}
    \caption{Snapshots of $\rho$ for the KHI test. The top panel shows the vortex formed with \bam\ at $t=10.9$ and the bottom panel shows the vortex formed with \sacra\ at $t = 10.0$.}
    \label{fig:KHI}
\end{figure}

The last two-dimensional special relativistic magneto-hydrodynamics test we consider in this work addresses the KHI as proposed in \cite{Mignone:2008ii,Bucciantini:2006jg}. For this test, we are focusing the comparison on the \bam\ and \sacra\ codes. We use the same setup as in \cite{Kiuchi:2022ubj} and specify a $\tanh$-shaped velocity profile with:
\begin{equation}
    v^x = -v_{\rm sh} \tanh \left(y/a\right).
\end{equation}
Here, $v_{\rm sh}$ determines the shear velocity and $a$ the thickness of the shear layer. As in \cite{Kiuchi:2022ubj}, we set $v_{\rm sh} = 0.25$ and $a = 0.02$. Further, we initialize a uniform density of $\rho = 1$ and pressure of $p=20$, and apply an ideal-gas EOS with $\Gamma = 4/3$. The components of the magnetic field are $B^x = \sqrt{2\sigma p}$ and $B^y = B^z = 0$. We set $\sigma = 0.01$. \par 

To disturb the shear layer, movement along the $y$ direction is introduced by:
\begin{equation}
    v^y = A_0 v_{\rm sh} \sin \left(2 \pi x\right) \exp \left(-100 y^2\right),
\end{equation}
where $A_0 = 10^{-4}$ determines the strength of the perturbation. The $v^z$ component remains zero. \par 

Figure~\ref{fig:KHI} presents snapshots of the density $\rho$ for \bam\ at $t=10.9$ and for \sacra\ at $t=10.0$. In both cases, the grid consists of one refinement level with $100 \times 200$ points covering a range for $x \in \left[-0.5, 0.5\right]$ and $y \in \left[-1, 1\right]$. We use in \bam\ MP5 and in \sacra\ PPM reconstruction. Both codes demonstrate their ability to resolve the vortex formed in the KHI test. The observable structure is very similar, although the vortex forms later for \bam\ than for \sacra. We attribute this to the usage of different Riemann solver: \bam\ applies HLL and \sacra\ HLLD. A similar behavior for different solvers is also found in \cite{Kiuchi:2022ubj}. 

\begin{figure}[t]
    \centering
    \includegraphics[width=\linewidth]{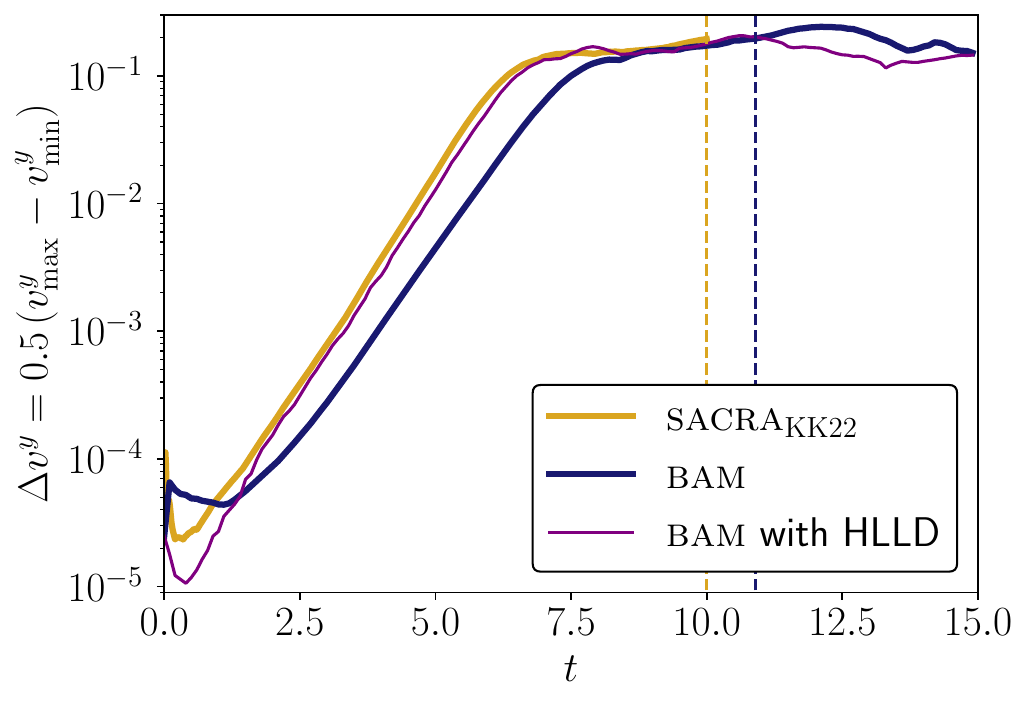}
    \caption{Time evolution for the perturbed velocity difference $\Delta v^y = \left(v^y_{\rm max} - v^y_{\rm min}\right) / 2$ of the KHI test for \bam\ and \sacra. The vertical dashed lines correspond respectively to times for the snapshots shown in Fig.~\ref{fig:KHI}.}
    \label{fig:KHI_Delta}
\end{figure}

\begin{figure*}[htp!]
    \centering
    \includegraphics[width=\linewidth]{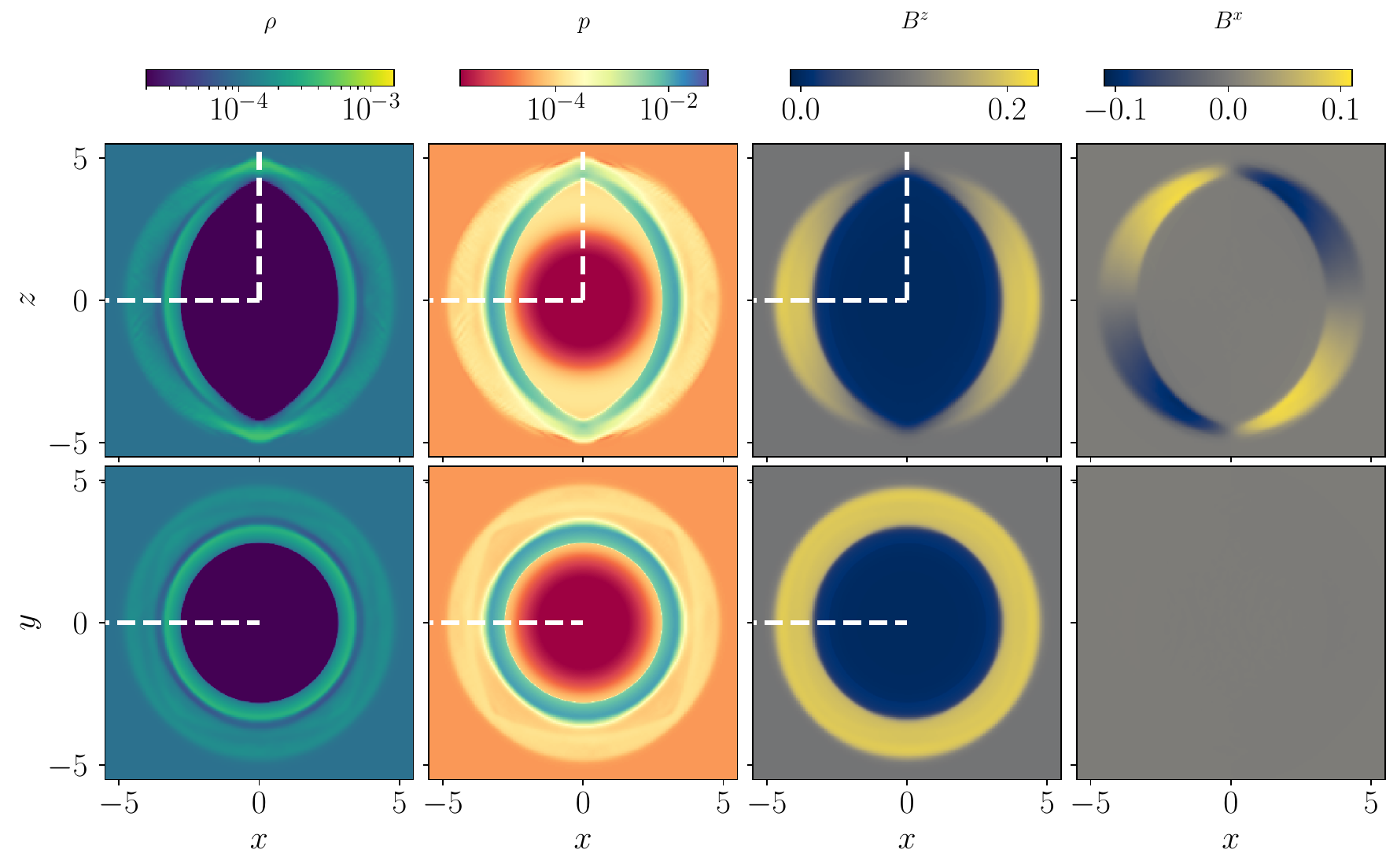}
    \caption{Snapshots of $\rho$, $p$, $B^z$, and $B^x$ for the spherical explosion test with \bam\ at the final time $t=4$ in the $x$-$z$ plane (upper panels) and $x$-$y$ plane (lower panels). The white dashed lines mark the one-dimensional profiles shown in Fig.~\ref{fig:SE1D}.}
    \label{fig:sphericalexplosion}
\end{figure*}

To be able to compare the simulations quantitatively, the perturbed velocity difference $\Delta v^y = 0.5 \left(v^y_{\rm max}-v^y_{\rm min}\right)$ is shown as a function of time in Fig.~\ref{fig:KHI_Delta}. We show the results for \bam\ additionally for using the HLLD solver that is still under development. As in \cite{Kiuchi:2022ubj}, $\Delta v^y$ grows exponentially until nonlinear saturation is reached. The growth rate is higher and nonlinear saturation is reached faster for \sacra\ and \bam\ using HLLD than for \bam\ using HLL, whereas the saturation amplitude is about the same for all cases. This demonstrates that the exponential growth for the KHI test depends strongly on the Riemann solver used, while the nonlinear saturation is only weakly dependent.

\subsection{Three-Dimensional Tests}

\subsubsection{Spherical Explosion}
\label{subsubsec:SE}

\begin{figure}[htp]
    \centering
    \includegraphics[width=0.9\linewidth]{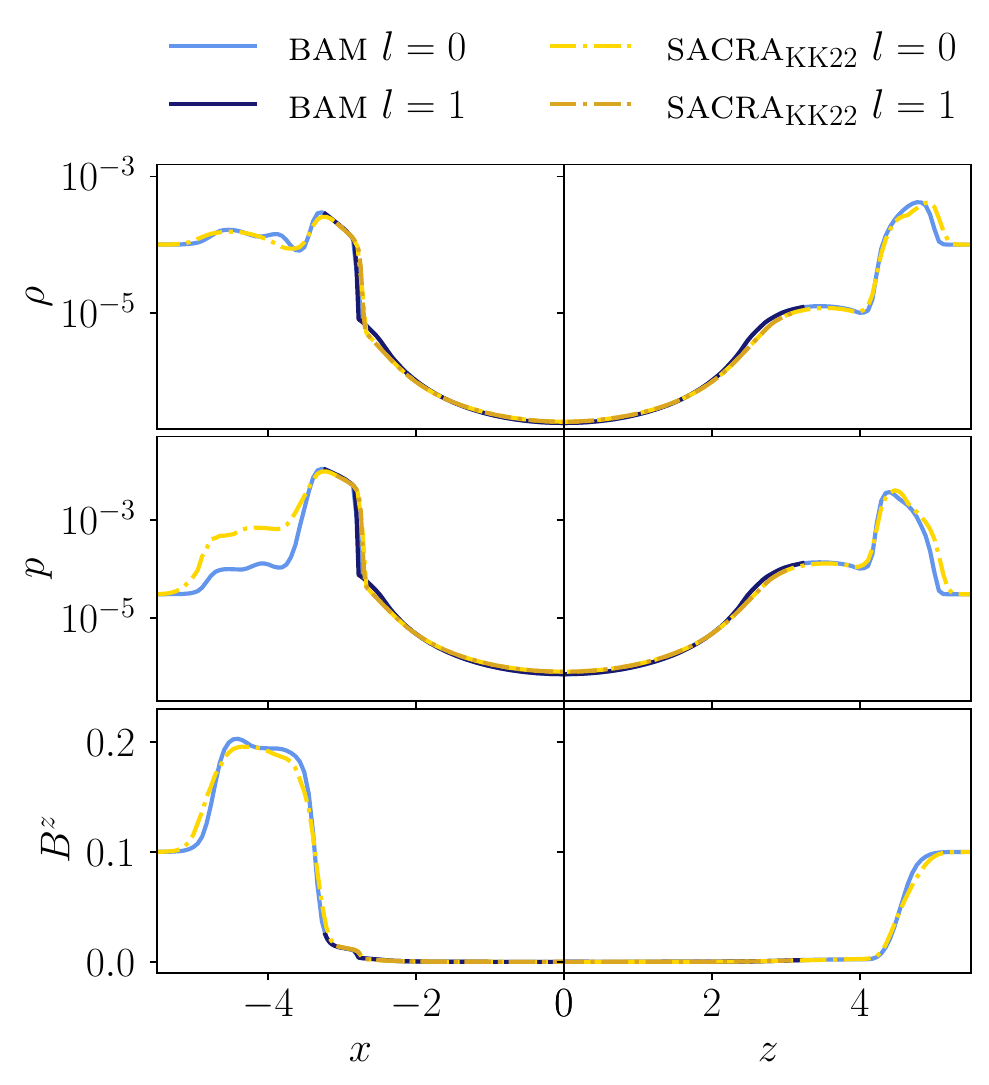}
    \caption{One-dimensional cuts along negative $x$- and positive $z$-axes of the spherical explosion test with \bam\ and \sacra. We show profiles for $\rho$, $p$, and $B^z$ on the refinement levels of $l=0$ and $l=1$.}
    \label{fig:SE1D}
\end{figure}

As a demanding three-dimensional shock problem for GRMHD codes, we perform the spherical explosion test with \bam\ and \sacra. We follow the description in \cite{Cipolletta:2019geh}, which corresponds to a three-dimensional version of the cylindrical explosion test in Sec.~\ref{subsubsec:CE}, where the cylindrical radius is substituted for a spherical radius: A dense medium with $\rho_{\rm in } = 10^{-2}$ and $p_{\rm in } = 1.0$ is set inside a spherical radius of $r_{\rm in} = 0.8$. The ambient medium outside of $r_{\rm out} = 1$ has $\rho_{\rm out } = 10^{-4}$ and $p_{\rm out } = 3 \times 10^{-5}$. In the transition region between $r_{\rm in} = 0.8$ and $r_{\rm out}=1$, we set:

\begin{align}
    \rho &= \exp \left( \frac{ \left(r_{\rm out} - r \right) \ln \rho_{\rm in} + \left(r - r_{\rm in} \right) \ln \rho_{\rm out} }{r_{\rm out}-r_{\rm in}} \right), \\
    p &= \exp \left( \frac{ \left(r_{\rm out} - r \right) \ln p_{\rm in} + \left(r - r_{\rm in} \right) \ln p_{\rm out} }{r_{\rm out}-r_{\rm in}} \right).
\end{align}
We apply again an ideal-gas EOS with $\Gamma = 4/3$ and initialize a uniform magnetic field with $B^x = B^y = 0.0$ and $B^z = 0.1$. The results obtained with \bam\ are shown in Fig.~\ref{fig:sphericalexplosion} at the final time $t=4$. For this test, we apply CENO3 reconstruction and use a grid with two refinement levels and $200 \times 200 \times 200$~points ranging from $\left[-6,6\right]$ in each direction. The grid spacing is accordingly $\Delta x = 0.06$ on $l=0$ and $\Delta x = 0.03$ on $l=1$. \par 

We compare in Fig.~\ref{fig:SE1D} our results for one-dimensional profiles along the $x$- and $z$-axes at the final time for $\rho$, $p$, and $B^x$. In this case, we focus on the comparison with \sacra\ and choose a grid with two refinement levels as already described above. We include in Fig.~\ref{fig:SE1D} lines for the profiles on level $l=0$ and $l=1$. Overall, the differences between \bam\ and \sacra\ are small: In the center, differences are of the order $10^{-8}$ for the rest-mass density, $10^{-7}$ for the pressure, and $10^{-5}$ for the magnetic field component $B^x$. The highest differences are found at the shock front of the $p$ profile along the $x$-axis. \sacra\ shows here larger values than \bam\ by a factor of almost ten. However, these larger differences are only outside the inner refinement level, where the resolution is quite low.\par 

\begin{figure}[t]
    \centering
    \includegraphics[width=0.9\linewidth]{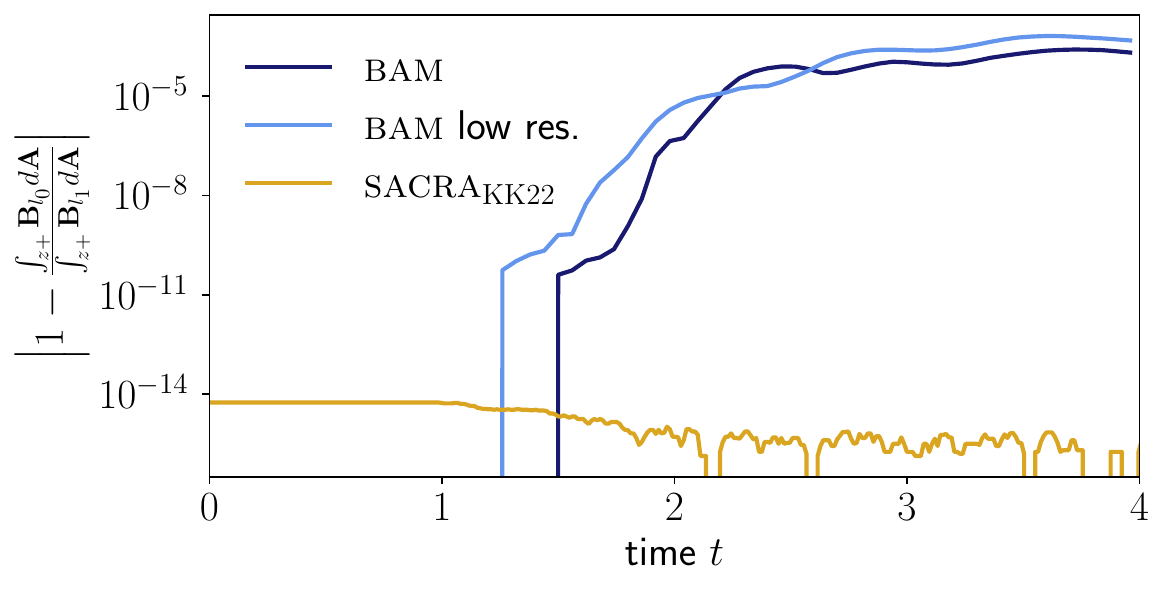}
    \caption{Magnetic flux conservation across refinement levels for the spherical explosion test with \bam, \sacra, and additionally with \bam\ for a lower resolution (labeled as `low res.'). Magnetic fluxes are computed as integral over the $z+$ refinement boundary, i.e., over the surface with $z=3$, $x \in [-3,3]$ and  $y \in [-3,3]$. We show the relative difference between the flux computed on $l=0$ and $l=1$. We use here a first-order Savitzky-Golay filter with the window length of $5$ samples for visualization purposes.}
    \label{fig:SE_flux}
\end{figure}

This test also enables us to access the conservation of the magnetic flux across refinement levels. By integrating the magnetic flux over the refinement boundary on levels $l=1$ and $l=0$, we can compare the magnetic flux violations. Figure~\ref{fig:SE_flux} shows the time evolution of the magnetic flux conservation across the positive $z$ refinement boundary, i.e., the area with $z=3$, $x \in [-3,3]$ and  $y \in [-3,3]$, and presents the relative difference between $\int_{z+} {\bf B}_{l_0} d{\bf A} $ and $\int_{z+} {\bf B}_{l_1} d{\bf A} $. We present results for \bam, \sacra, and additionally for \bam\ using a lower resolution with $\Delta x = 0.12$ on $l=0$ and $\Delta x = 0.06$ on $l=1$ (labeled as `low res.'). While both codes use methods to correct fluxes across refinement boundaries, it is evident that the magnetic flux is better conserved in the simulation with \sacra\ than in the simulations with \bam. For \sacra, the violations are of the order of $10^{-14}$, i.e., at round-off accuracy. For \bam, the difference between the fluxes is initially zero, but increases significantly once matter crosses the refinement boundary at $t \approx 1.5$, reaching orders of $10^{-5}$. Using \textit{divergence cleaning} in \bam, the evolution equation for the magnetic field contains additional source terms. Although \bam\ incorporates algorithms to correct the fluxes across refinement boundaries, the source terms are not matched and may marginally differ, resulting in slightly different values. Figure~\ref{fig:SE_flux} shows small improvements for increased resolution.

\section{Binary Neutron Star Merger Simulations}
\label{sec:BNSSimulations}

\begin{figure*}[htp]
    \centering
    \includegraphics[width=\linewidth]{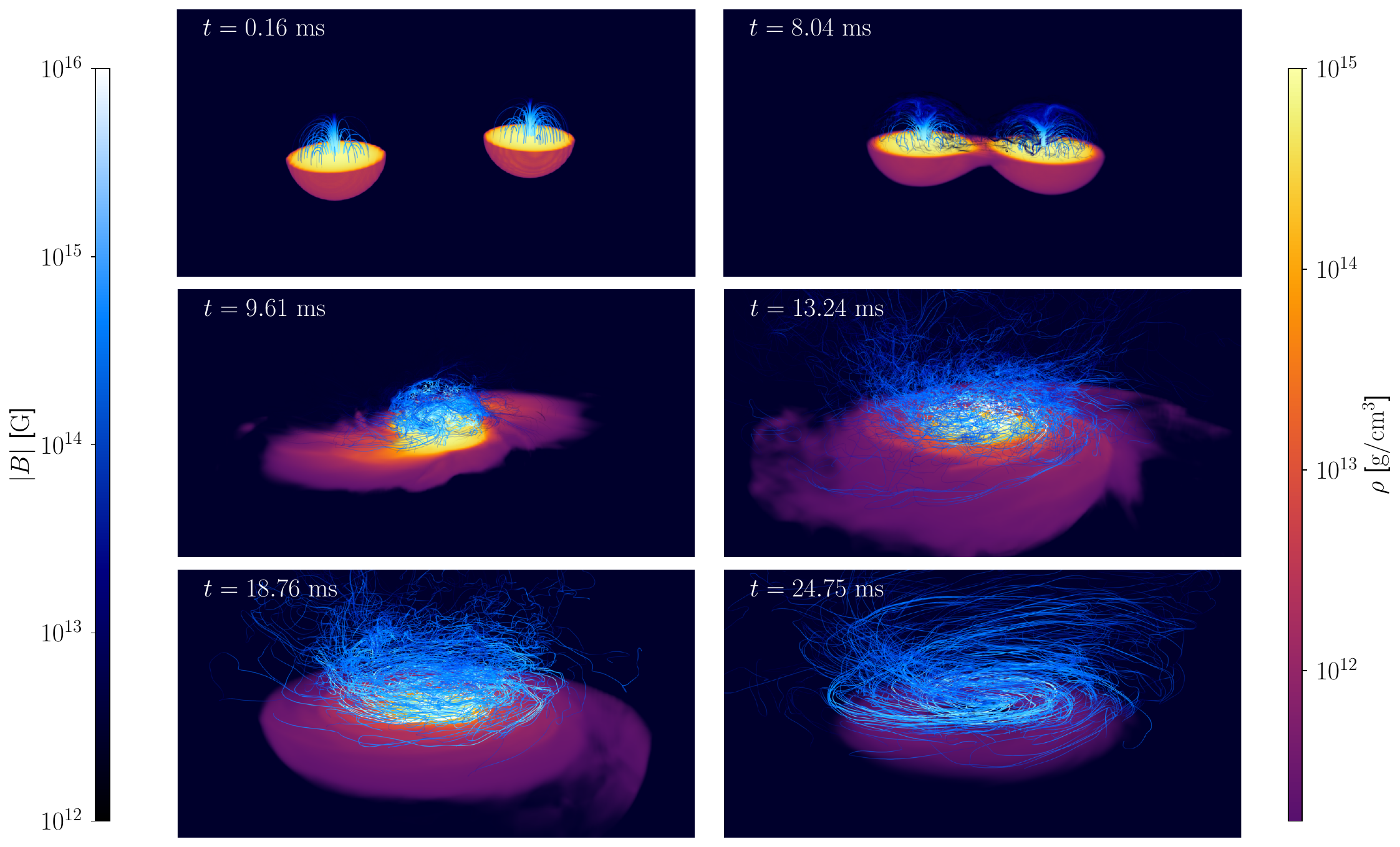}
    \caption{Three-dimensional snapshots of the BNS simulation using \bam\ for the highest resolution. In the lower half with a cut at $z=0$ the rest-mass density $\rho$ is shown, while the magnetic field lines are shown in blue in the upper half.}
    \label{fig:BNS3D}
\end{figure*}

\subsection{Configurations}

We perform simulations of BNS merger with magnetic fields and compare the simulation results with \sacra. The publicly-available, spectral code \textsc{fuka}~\cite{Papenfort:2021hod} is used to construct initial data for an equal-mass BNS system with gravitational masses $m_A = m_B = 1.35\ M_\odot$ and baryonic masses $m_{b,A} = m_{b,B} = 1.49\ M_\odot$. As \sacra\ has an implemented \textsc{fuka} reader, the same initial data is used to directly compare the evolution and results of the BNS simulations. We set an initial coordinate separation of $\sim 28.4$~km to ensure a short inspiral, focusing the simulation on the merger and post-merger phases. The initial ADM mass is $M_{\rm ADM} =2.673\ M_{\odot}$. For the EOS, a piece-wise polytropic fit of SLy~\citep{Douchin:2001sv} is applied following \cite{Read:2008iy} with four pieces, three for the NS core and one for the crust. In order to include thermal effects, the zero-temperature EOS is extended by a thermal pressure $P_{\rm th} = \left(\Gamma_{\rm th} - 1\right) \rho \epsilon_{\rm th}$ \cite{Bauswein:2010dn}, where $\epsilon_{\rm th}$ is the thermal part of the specific internal energy and we set $\Gamma_{\rm th} = 1.75$. We note that a slightly different value is used for the simulations with \sacra, namely $\Gamma_{\rm th} = 1.8$. \par

We initialise a poloidal magnetic field for each star by a vector potential:
\begin{align}
    A_x &= -\Tilde{y} A_b \max(p - p_{\rm cut},0)^2, \\
    A_y &=  \Tilde{x} A_b \max(p - p_{\rm cut},0)^2, \\
    A_z &=  0,
\end{align}
with $\Tilde{x} = x - x_{\rm NS}$ and $\Tilde{y} = y - y_{\rm NS}$ as distances from the center of the NS along $x$ and $y$, respectively. We set $A_b = 1000$ and $p_{\rm cut}=0.004 \times p_{\rm max}$, where $p_{\rm max}$ is the maximum pressure in the NS at $t=0$\,ms. In this way, we obtain a maximum magnetic field strength of the order $\sim 10^{15}$\,G inside the stars. \\

For the grid, we use in \bam\ eight refinement levels where the three outermost are non-moving and the five innermost are moving levels. The simulations are performed on three different resolutions: R1 with $n = 128$ and $n_{\rm mv} = 64$ grid points, R2 with $n = 256$ and $n_{\rm mv} = 128$ grid points, and R3 with $n = 512$ and $n_{\rm mv} = 256$ grid points in each direction. The grid spacing on the finest box covering the NS are $\Delta x_{\rm min} \approx 368.6$\,m for R1, $\Delta x_{\rm min} \approx 184.3$\,m for R2, and $\Delta x_{\rm min} \approx 92.1$\,m for R3. We use the HLL Riemann solver as described in Sec.~\ref{subsubsec:riemannsolver} and the reconstruction method as described in Sec.~\ref{subsubsec:reconstruction} with MP5 as high-order reconstruction and CENO3 as low-order reconstruction, i.e., we examine oscillations and fall back to CENO3 or further to linear reconstruction if necessary. We employ this only for low density regions: in the R1 and R2 simulations for $\rho < 100 \times \rho_{\rm atm}$ and in the R3 simulation for $\rho < 10^6 \times \rho_{\rm atm}$\footnote{At the highest resolution, we encountered issues when using MP5 due to larger oscillations leading to unphysical behavior of the matter outflow along the grid axes. This problem could be solved with a higher threshold for the rest-mass density using our scheme to detect enhanced oscillations and maintain the positivity of the reconstructed density and pressure.}.\par

\sacra\ uses a similar grid configuration with ten refinement levels and the same resolutions R1, R2 and R3 as in \bam. However, the box sizes of the refinement levels are larger as \sacra\ uses a fixed mesh refinement (FMR). The half size of the finest level is set to $L_{\rm fin} =35.5~{\rm km}$. For the inner part of the finest refinement level, a constant atmosphere is set with $\rho_{\rm atm} = 10^3\ \rm g/cm^3$. Outside a power-law profile is applied with $\rho_{\rm atm} = 10^3 \left(L_{\rm fin}/r\right)^3 \ \rm g/cm^3$. \sacra\ introduces a restriction for high magnetic-pressure regions with $b^2/\rho \geq 10^3$ by artificially suppressing the specific momentum. The reason is that the numerical accuracy is not good in those regions and, as a result, the primitive recovery procedure often fails. However, in the current work, in which the post-merger evolution is not followed over a long timescale, we do not find such a highly magnetized region. As in Sec.~\ref{sec:FirstTests}, \sacra\ applies the HLLD solver with third-order PPM reconstruction.

\subsection{Evolution and Magnetic Field Amplification}

\begin{figure}[t]
    \centering
    \includegraphics[width=\linewidth]{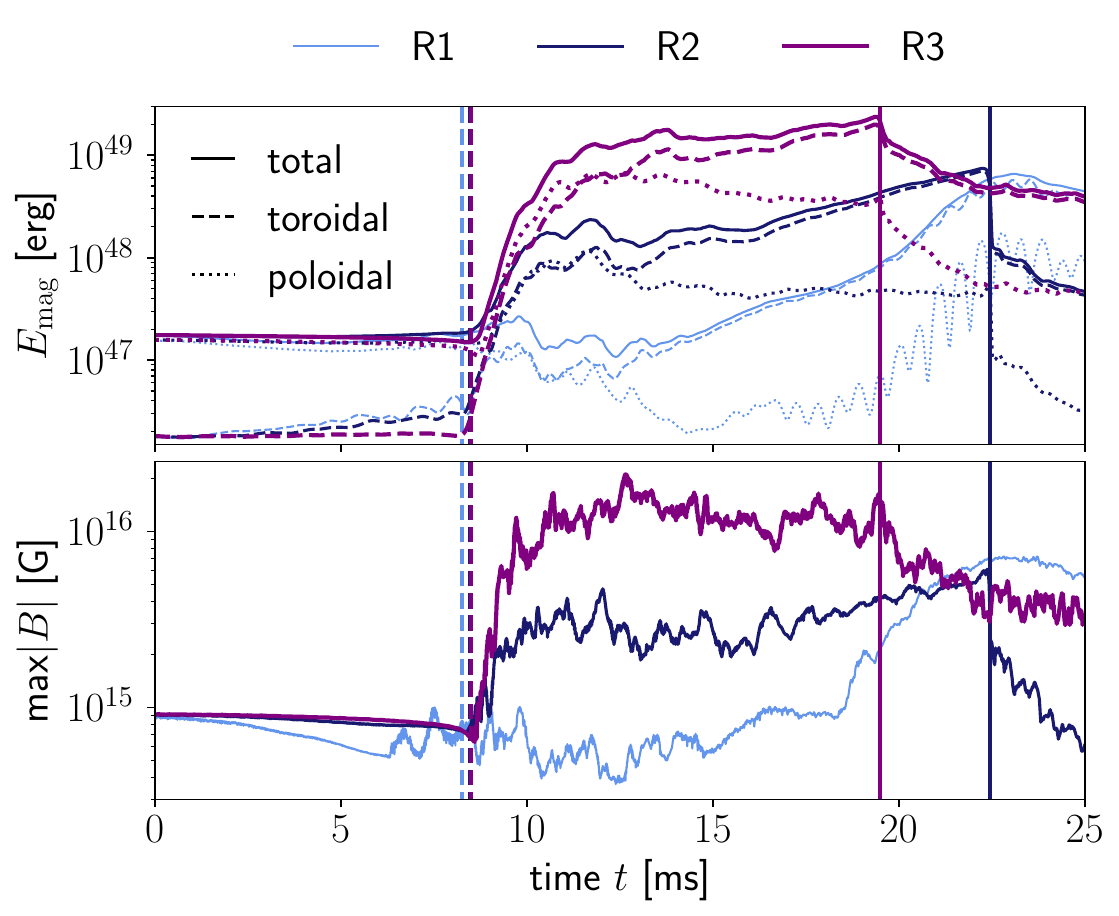}
    \caption{Time evolution of magnetic energy (upper panel) and maximum magnetic field strength (lower panel) for the BNS simulations with \bam\ with resolution R1, R2, and R3. The dashed lines show the toroidal part and dotted lines poloidal part of the magnetic energy. The magnetic energy is extracted from refinement level $l=1$ and the maximum magnetic field strength from $l=6$. The vertical lines show the respective merger time (dashed) and collapse time (solid).}
    \label{fig:BNS-res}
\end{figure}

\begin{figure*}[t]
    \centering
    \includegraphics[width=\linewidth]{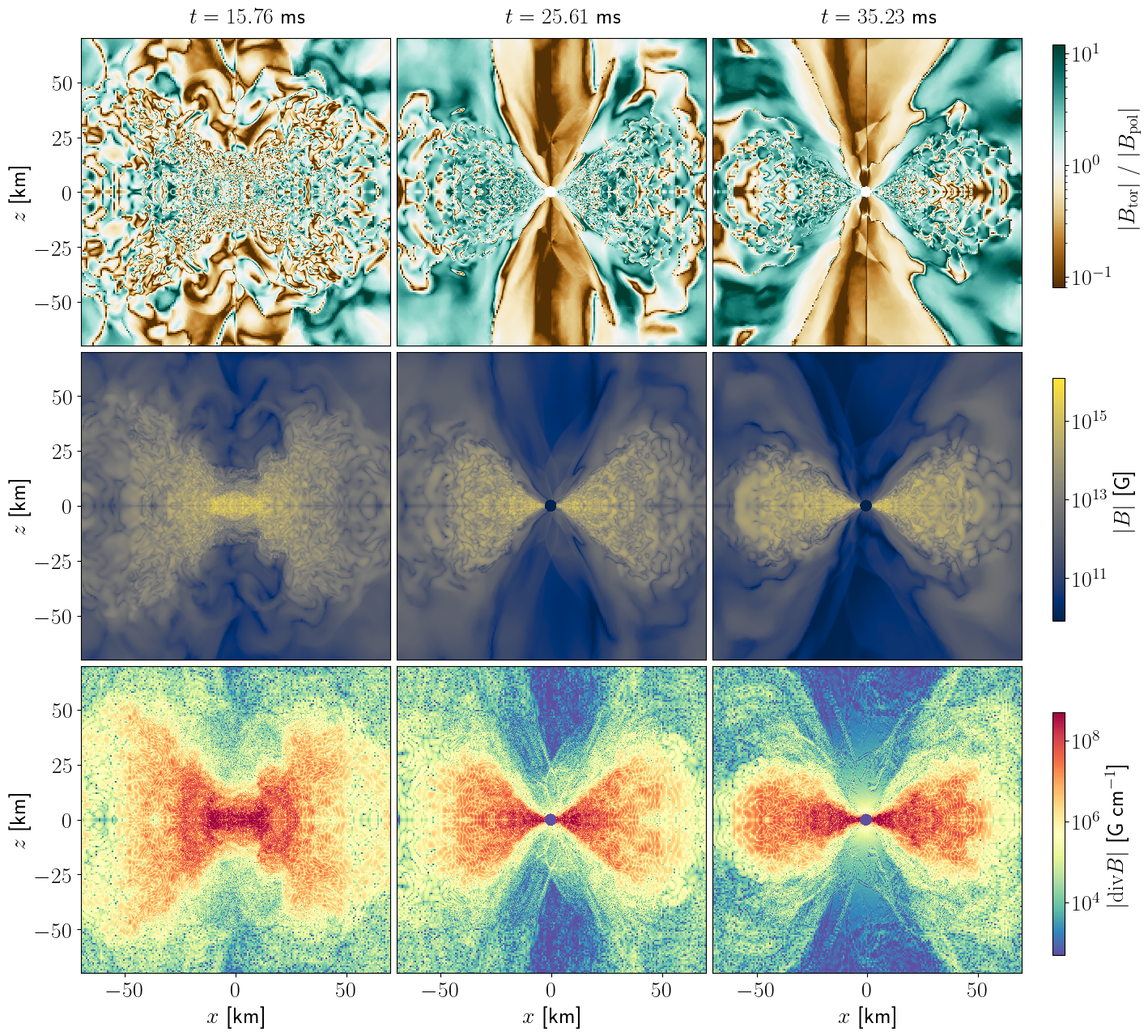}
    \caption{Magnetic field snapshots on the $x$-$z$ plane after the merger of the BNS simulation with \bam\ for the highest resolution. We show the ratio between the toroidal and poloidal component in the upper row, the absolute magnetic field strength in the middle row, and the divergence of the magnetic field in the bottom row.}
    \label{fig:BNS-torvspol}
\end{figure*}

The evolved BNS system merges after about four orbits and forms a hypermassive neutron star (HMNS), collapsing into a BH within a few tens of milliseconds. The lifetime of the HMNS varies for different resolutions because the HMNS just prior to the collapse is marginally stable, and hence, small perturbations including numerical errors can trigger the collapse. In Fig.~\ref{fig:BNS3D}, we present snapshots of the rest-mass density and the magnetic field lines for the simulation with the highest resolution. The snapshots show the two NSs at the beginning of the simulation, at the merger (at about $9.6\ \rm ms$), and the remnant before and after the collapse. The initial magnetic dipole field inside the NSs is clearly visible, as well as the winding of the magnetic field lines during the merger, forming a helicoidal structure. \par

We show the time evolution for the magnetic energy and maximum magnetic field strength in Fig.~\ref{fig:BNS-res} for the R1, R2, and R3 simulations with \bam. The magnetic energy is defined by:
\begin{equation}
    E_{\rm mag} = \frac{1}{2} \int u^0 \sqrt{-g} b^2 d^3x.
\end{equation}
Additionally, the toroidal and poloidal contribution are shown. We note that they are defined with respect to the coordinate center, which is only approximately correct after the merger assuming the remnant stays in the center. During the merger, the magnetic field is amplified. We observe here a strong resolution dependency: For R1, there is almost no amplification during the merger itself. Only on later time scales of about $10\ \rm ms$ after the merger the magnetic energy and field strength rise. For R2, the maximum magnetic-field strength doubles, and for R3, it increases tenfold, leading to energies up to $10^{49}\ \rm erg$. The amplification is triggered by a variety of instabilities. For example, the shear layer between the two NS induces a KHI during the merger, which forms vortices that twist the magnetic field and lead to exponential growth, e.g.,~\cite{Kiuchi:2015sga, Kiuchi:2017zzg}. Another mechanism is driven by large-scale differential rotation in the post-merger, which winds up the magnetic field lines. While the magnetic winding can already be resolved with lower resolutions, capturing the KHI and also the MRI requires higher resolutions. In order to fully capture these instabilities and the associated magneto-hydrodynamic turbulence, even higher resolutions are required as illustrated in~\cite{Kiuchi:2015sga, Kiuchi:2017zzg} (see also \cite{Kiuchi:2024lpx}). \par

Although the magnetic field is initially purely poloidal, the toroidal component becomes dominant after the merger as shown in Fig.~\ref{fig:BNS-res}. For a qualitative analysis of the magnetic field in the remnant system, Fig.~\ref{fig:BNS-torvspol} presents snapshots in the $x$-$z$ plane after the merger. We show the magnetic field strength and the ratio between the toroidal and poloidal component before and after the collapse for the R3 simulation. Before the collapse, the magnetic field in the polar region is already predominantly poloidal. This becomes even clearer after the collapse. The disk of the remnant has rather toroidal magnetic fields. After the collapse, the magnetic field in the polar region decreases from an order of magnitude of $10^{12}\ \rm G$ to an order of magnitude of $10^{11}\ \rm G$. The magnetic field is strongest inside the disk with up to $10^{16}\ \rm G$. \par 

Additionally, we show in Fig.~\ref{fig:BNS-torvspol} the divergence of the magnetic field. As already mentioned, a limitation of the implemented \textit{divergence cleaning} method is that the divergence of the magnetic field is not exactly zero and cannot prevent the formation of magnetic monopoles. To assess how large the divergence actually is, we present maps of the remnant after the merger. It is largest in the interior of the disk, where the magnetic field is also strongest, with values in the order of $10^8\ \rm G / cm$ and smallest in the polar regions with values in the order of $10^3\ \rm G / cm$. \par 

\begin{figure}[t]
    \centering
    \includegraphics[width=\linewidth]{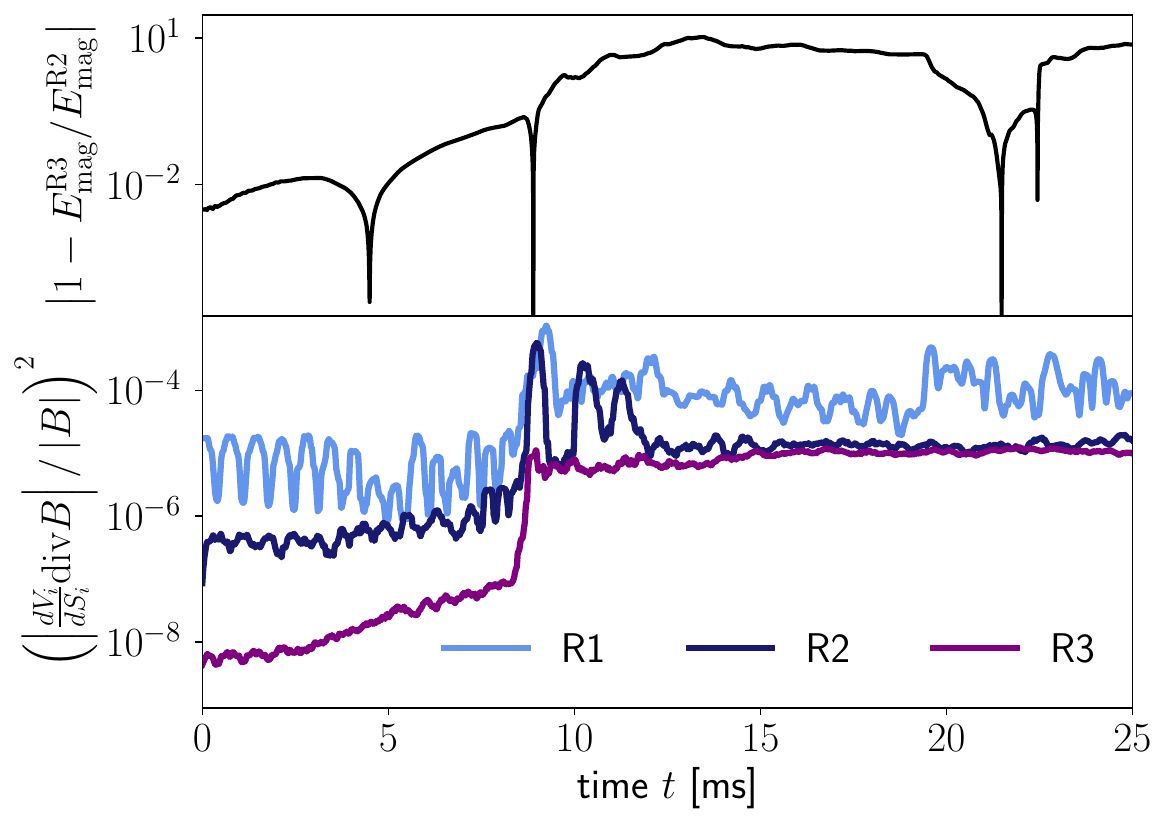}
    \caption{Time evolution of the relative error on the magnetic energy with \bam; cf. main text. \textit{Upper panel}: Relative difference on the magnetic energy between R2 and R3 simulation. \textit{Lower panel}: Estimated relative error caused by the violation of the magnetic monopole conservation. The divergence is multiplied by the volume of a grid cell $dV_i$ and divided by its surface area $dS_i$ to obtain the error on the magnetic field. Normalizing this quantity by the magnetic field strength and squaring it provides an estimate of the relative error in magnetic energy caused by the divergence. The quantities are extracted from level $l=1$. We use in the lower panel a first-order Savitzky-Golay filter with the window length of $50$ samples for visualization purposes.}
    \label{fig:BNS-div}
\end{figure}

One important point is to compare errors introduced by the magnetic field divergence violation with other uncertainties, e.g., due to finite resolution, flux computation, or shock treatments. Following Gauss's law, we can translate the divergence into an error on the magnetic field within a grid cell by multiplying with its volume $dV_i$ and dividing by its surface area $dS_i$. With $E_{\rm mag} \propto B^2$, we obtain an estimate of the relative error on the magnetic energy by normalizing with the magnetic field strength and squaring the quantity. In Fig.~\ref{fig:BNS-div}, we compare this relative error of the simulations using R1, R2, and R3 with the relative difference of the magnetic energy between resolution R2 and R3. The error in the magnetic energy due to the magnetic field divergence increases during the merger. Overall, the relative error is between $10^{-8}$ and $10^{-4}$ and decreases with higher resolution. The relative difference between the magnetic energy at R2 and R3 resolution is about $10^{-2}$ and increases to $10$ after the merger. We note that, in agreement with other GRMHD simulations shown in previous studies, even our highest resolution is well outside the convergence regime, explaining the large differences here. Nevertheless, this indicates that the errors due to the divergence of the magnetic field when using the \textit{divergence cleaning} scheme are minor if not negligible compared to the errors caused by other resolution-dependent effects. 

\subsection{Comparison}
\label{subsec:BNScomparison}

\begin{figure}[t]
    \centering
    \includegraphics[width=\linewidth]{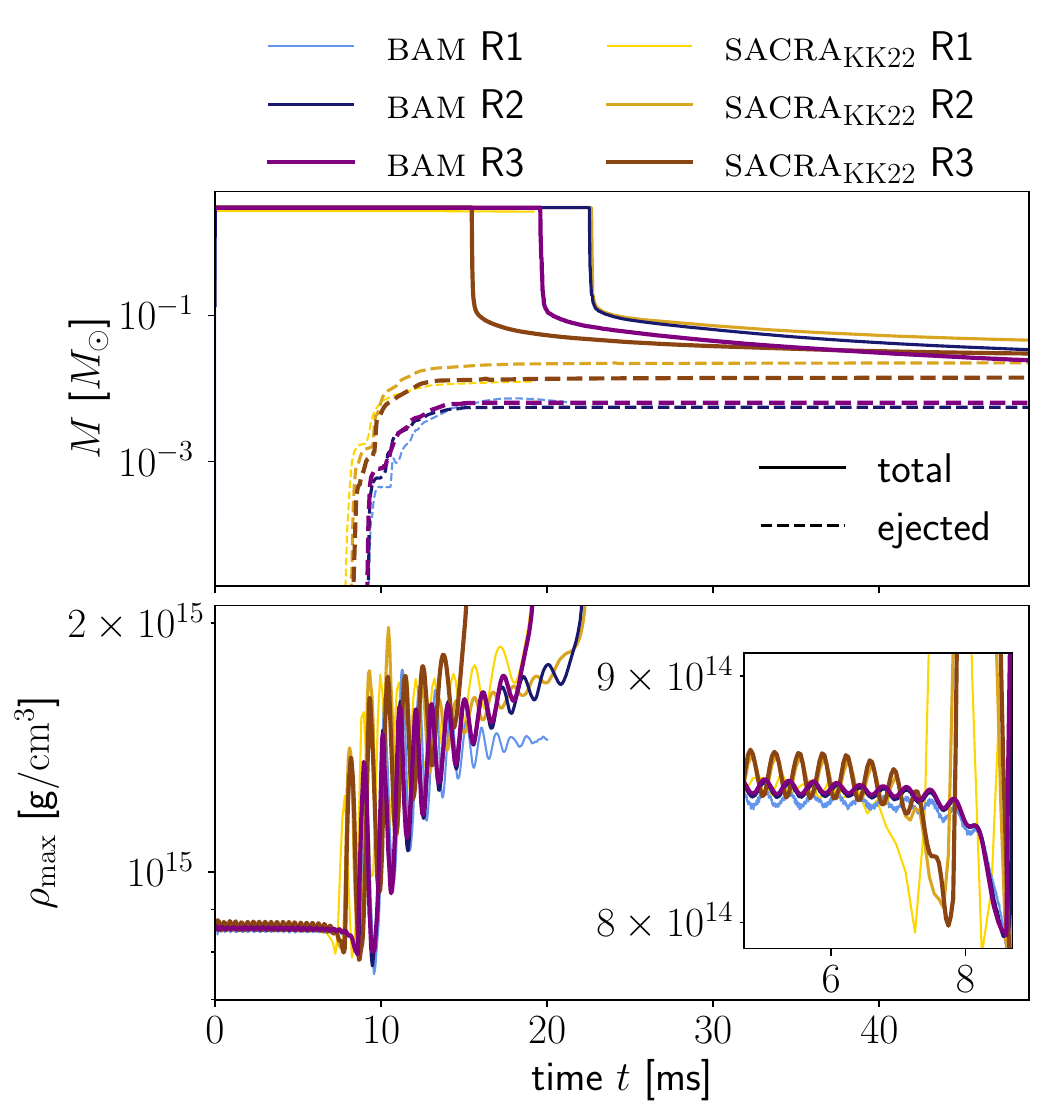}
    \caption{Comparison of total mass, ejecta mass, and central rest-mass density in BNS simulations with \bam\ and \sacra\ for resolutions R1, R2, and R3. For the \bam\ results, total and ejecta mass are extracted from refinement level $l=1$ and the central rest-mass density from $l=6$.}
    \label{fig:BNScomp1}
\end{figure}

\begin{figure}[t]
    \centering
    \includegraphics[width=\linewidth]{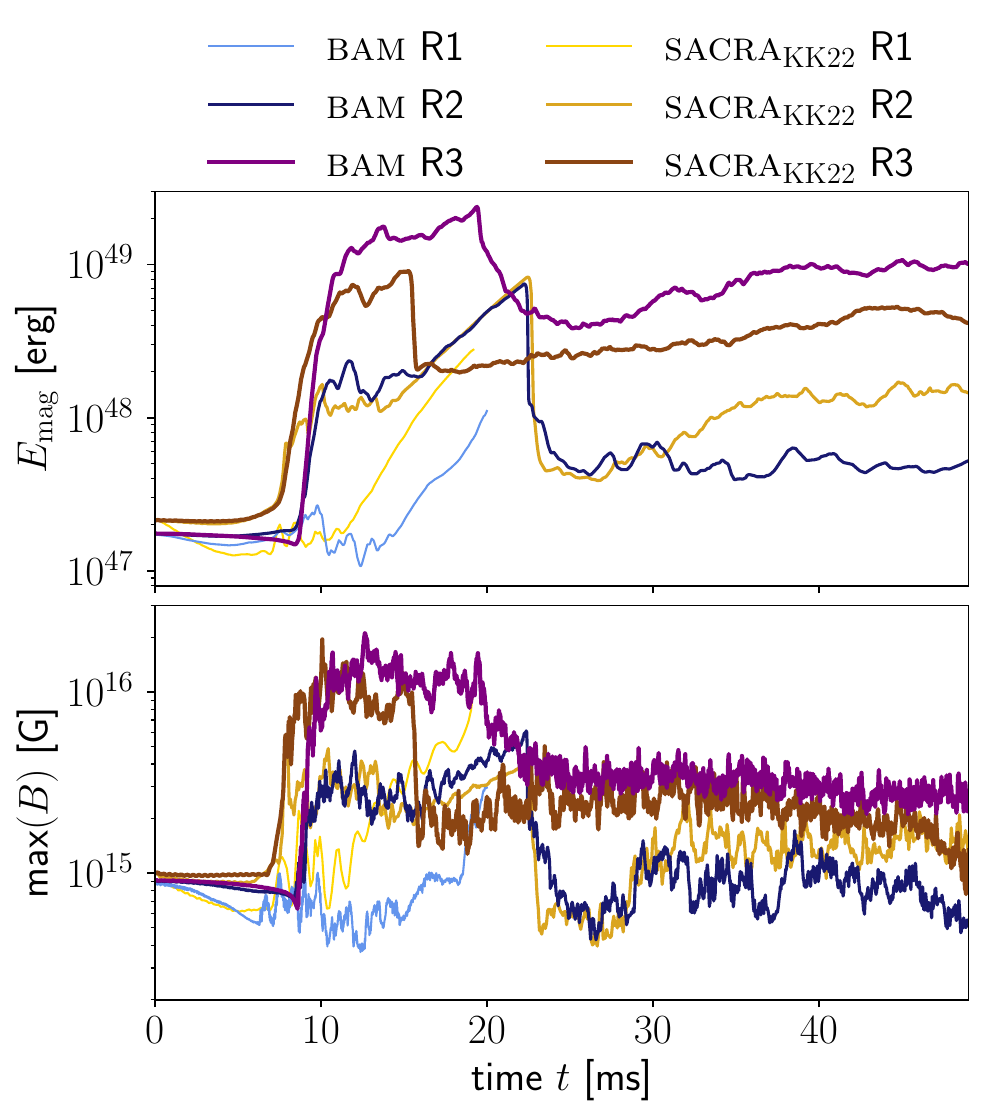}
    \caption{Comparison of magnetic energy and maximum magnetic field strength in BNS simulations with \bam\ and \sacra\ for resolutions R1, R2, and R3. For the \bam\ results, the magnetic energy is extracted from refinement level $l=1$ and the maximum magnetic field strength from $l=6$.}
    \label{fig:BNScomp2}
\end{figure}

Finally, we compare the results of the BNS simulations performed with \bam\ and \sacra\ in Figs.~\ref{fig:BNScomp1} and \ref{fig:BNScomp2}. The time evolution of the total baryonic mass, ejecta mass, and the central density inside the NSs is shown in Fig.~\ref{fig:BNScomp1} and the magnetic energy and maximum field in Fig.~\ref{fig:BNScomp2} for both codes at resolution R1 to R3. We use the geodesic criterion for the ejecta. Accordingly, matter is considered unbound if the condition $u_t < -1$ with a positive radial velocity $v_r >0$ is satisfied. We find that the merger time in the simulations with \sacra\ is about $1.4\ \rm ms$ earlier than in the simulations with \bam. This is due to the different gauge conditions used in \sacra. We discuss this in more details in Appendix~\ref{app:hydro}. 

The ejecta mass is considerably larger in the simulations performed with \sacra\ than in the ones performed with \bam, i.e., by a factor of $\sim 1.9$ for resolution R1, $\sim 4.1$ for resolution R2, and $\sim 2.4$ for resolution R3. Thereby, the variation for different resolutions is greater for \sacra\ ranging between $0.015\ M_\odot$ and $0.022\ M_\odot$ than for \bam\ ranging between $0.0055\ M_\odot$ and $0.0064\ M_\odot$. A possible reason for the larger variations could be attributed to using different reconstruction schemes with different orders in \sacra\ and \bam. Another reason that could explain the strong differences in ejecta mass is the treatment of the vacuum region. As described in Sec.~\ref{subsubsec:atmosphere}, \bam\ uses an artificial atmosphere with $f_{\rm atm} = 10^{-11}$, resulting in atmosphere densities of the order of $10^4\ \rm g/cm^3$, while \sacra\ uses lower atmospheric values of at most $10^3\ \rm g/cm^3$, which further decreases with increasing radius. This can lead to differences in the ejecta as the outward flowing matter travels more freely. The lower atmosphere densities can therefore explain the higher ejecta masses in the simulations with \sacra. \par 

\begin{table}[t]
\caption{Properties of the remnant system, from left to right: code name, resolution, ejecta mass, disk mass, BH mass, and total energy radiated by GWs $E_{\rm GW}$. The values are extracted at the end of the simulation at $t \approx 50\ \rm ms$.}
\label{tab:remnant}
\begin{tabular}{c c||c|c|c|c}
\toprule 
code  & res.  & $M_{\rm BH} [M_{\odot}]$ & $M_{\rm disk} [M_{\odot}]$ & $M_{\rm eje} [M_{\odot}]$ & $E_{\rm GW} [M_{\odot}]$  \\ 
\hline \hline
\bam      & R2 &  2.512 &  0.0264 &  0.0055 &  0.0787 \\
\bam      & R3 &  2.543 &  0.0177 &  0.0064 &  0.0871 \\
\hline
\sacra    & R2 &  2.560 &  0.0213 &  0.0225 &  0.0571 \\
\sacra    & R3 &  2.569 &  0.0158 &  0.0152 &  0.0705 \\
\bottomrule
\end{tabular}
\end{table}

The values for the final ejecta mass, disk mass, remnant BH mass, and energy radiated by GWs evaluated at the end of the simulations at $t \approx 50\ \rm ms$ are listed in Tab.~\ref{tab:Balsara}. Since resolution R1 is too low to give reliable results after the BH formation, we stopped the simulation for \sacra\ after the collapse. For \bam, there was no BH formation within the simulation period for R1. Results for R1 simulations are therefore not listed in Tab.~\ref{tab:Balsara}. Indeed, the remnant BH masses and disk masses are more similar for both codes. Comparing the values for R3 resolution, \sacra\ predicts a slightly larger BH mass of $\sim 2.57\ M_\odot$ than \bam\ with $\sim 2.54\ M_\odot$. On the other hand, the disk masses in the simulations with \bam\ are greater than the one with \sacra\ by about $\sim 0.005\ M_\odot$ for R2 and $\sim 0.002\ M_\odot$ for R3. In order to assess the error size in the conservation of the energy, we compute $\Delta E:=M_{\rm ADM} - M_{\rm BH} - M_{\rm disk} - M_{\rm eje}-E_{\rm GW}$. For R2 and R3 in \bam, we get respectively $\Delta E \approx 0.0504 \ M_{\odot}$ and $0.0188 \ M_{\odot}$, while for R2 and R3 in \sacra, we get $\Delta E \approx 0.0121 \ M_{\odot}$ and $0.0025 \ M_{\odot}$. The values of $\Delta E$ for R3 are much smaller than in R2 for both codes, showing good convergent behavior. The error size for R3 in \sacra\ is slightly smaller than that in \bam, which we suggest comes from the different settings of the artificial atmosphere influencing the measured $M_{\rm eje}$.

\begin{figure}[t]
    \centering
    \includegraphics[width=\linewidth]{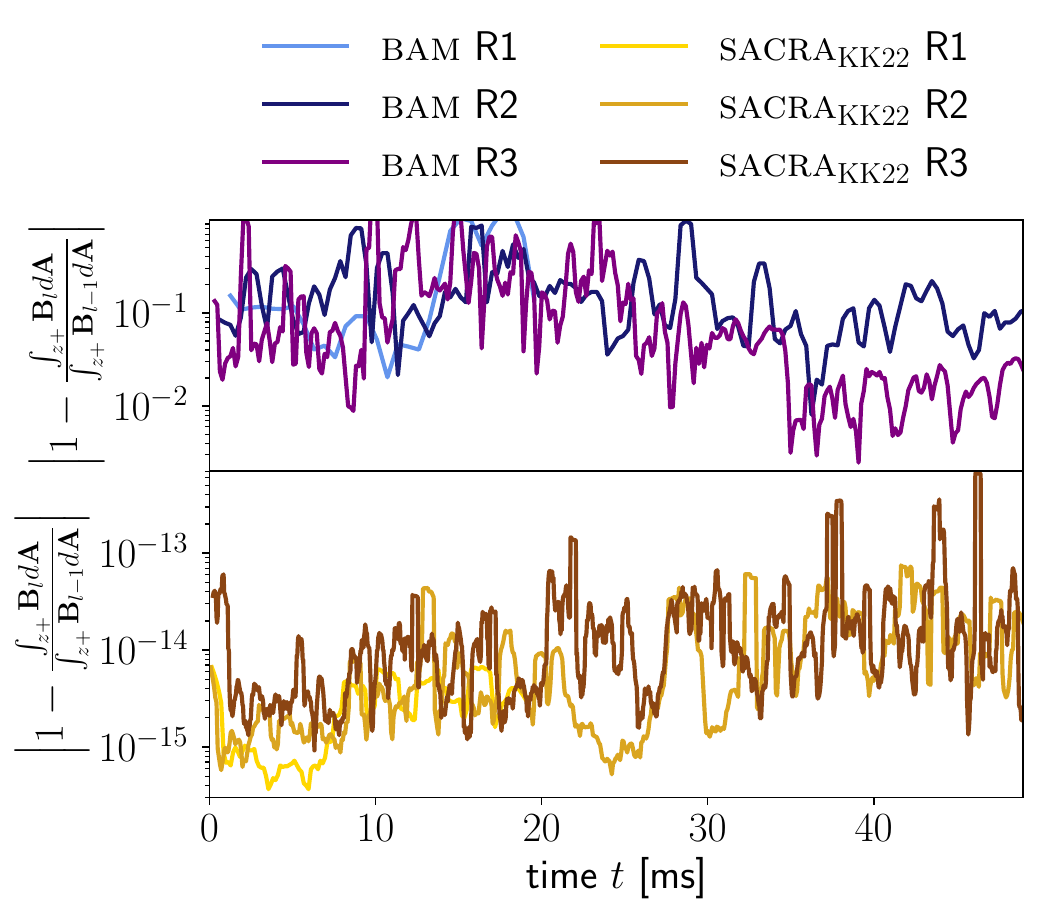}
    \caption{Conservation of the magnetic flux across refinement boundaries in BNS simulations with \bam\ and \sacra\ for resolutions R1, R2, and R3. Magnetic fluxes are computed as integral over the refinement boundary along the positive $z$ direction: for \sacra\ at the boundary of level $l=10$ and for \bam\ at $l=6$. For visualization purposes, we use a first-order Savitzky-Golay filter with window length of 10 samples for \sacra\ simulations and 3 samples for \bam\ simulations.}
    \label{fig:BNSmagflux}
\end{figure}

Considering the violation of the divergence constraint for the magnetic field, we find larger violations for \bam\ than for the \sacra\ simulations, caused by the employed evolution schemes. The norm of the magnetic field divergences normalized to the norm of the magnetic field strength lies for \bam\ between $10^{-9} \ \rm cm^{-1}$ and $5 \times 10^{-8} \ \rm cm^{-1}$ and for \sacra\ between $10^{-19} \ \rm cm^{-1}$ and $2 \times 10^{-18} \ \rm cm^{-1}$. \par

In addition, we analyze the conservation of the magnetic flux across refinement boundaries. For this purpose, we integrate the magnetic field across the refinement boundaries at level $l$ and its adjacent coarser level $l-1$, and compute the relative difference of the magnetic flux similarly as in the spherical explosion test. We show the results along the positive $z$ direction for \sacra\ at the refinement boundary of level $l=10$ and for \bam\ at level $l=6$. The comparison here is not straightforward since the refinement boxes for \sacra's FMR are much larger than for \bam. The refinement boundary of the finest level of \sacra\ lies between the refinement boundaries of $l=6$ and $l=5$ of \bam, whereby the boxes move dynamically within \bam. Nevertheless, it is visible that the relative differences for \bam\ are orders of magnitude larger than for \sacra: for \bam\ at orders of  $10^{-2}$ and $10^{-1}$, whereas for \sacra\ at orders of $10^{-15}$ to $10^{-14}$, i.e., machine accuracy. \par

However, for comparing the magnetic energy and the maximum field strength in Fig.~\ref{fig:BNScomp2}, both codes predict overall similar values for the respective resolutions. In the R1 simulations, there is almost no amplification during the merger in both cases. The magnetic field only increases after the merger, which happens slightly earlier for \sacra\ than for \bam, but with a similar slope in the magnetic energy. With R2, both codes show an amplification of the magnetic field during the merger up to $10^{48}\ \rm erg$. Despite the different merger time, the collapse time coincides at this resolution for both codes, and the lines for the magnetic energy almost overlap between $17\ \rm ms$ and $25\ \rm ms$. Thereafter, \sacra\ predicts slightly higher energies than \bam. In the highest-resolution simulation, both codes reach a maximum magnetic field strength of $\sim 10^{16}\ \rm G$ shortly after the merger. The simulation performed with \bam\ reaches higher magnetic energies than \sacra\ by a factor of two, which could be caused by the higher-order reconstruction in \bam\ that possibly resolves small-scale effects slightly better. The collapse time is a few milliseconds earlier with \sacra\ than with \bam. In general, both codes predict similar magnetic energies and field strengths at the end of the simulation.

\section{Conclusions}
\label{sec:Conclusions}

We successfully extended the infrastructure of the \bam\ code for ideal GRMHD simulations employing a hyperbolic \textit{divergence cleaning} scheme. Additionally, we compared results for well-known special-relativistic tests with established GRMHD codes: \spritz, \gramx, and \sacra. 
Overall, the tests showed good agreement between all codes. In the one-dimensional Balsara test, we observed only minor differences at the shock fronts, which can be attributed to the fact that the reconstruction methods used in the individual codes are either more diffuse or more oscillatory.
Similarly, we found minor differences in the two-dimensional cylindrical explosion and magnetic rotor tests, which we attribute to different reconstruction schemes for the fluid variables. The largest differences occur for \sacra, whereby higher resolution converges to the same results. The \sacra\ code is the only one using the HLLD Riemann solver for these tests. The other codes use the HLL solver, which could explain the larger differences here.
We also compare results for the Kelvin-Helmholz instability test between \bam\ and \sacra. Both codes show that they are able to capture the vortex. In \sacra, the vortex forms earlier than in \bam\ and the perturbation grows faster. However, we show that we achieve the same growth rate in \bam\ when using the same Riemann solver as in \sacra.
In the three-dimensional spherical explosion test, we compare the conservation of magnetic flux over a refinement boundary for \bam\ and \sacra. Both codes apply schemes to conserve the flux. Still, \sacra\ proves to be superior in magnetic flux conservation. Applying \textit{divergence cleaning} in \bam\ leads to additional source terms in the evolution equation for the magnetic field, which are not matched and could explain the worse performance. \par

We performed first BNS simulations with our new \bam\ implementation and compare our results with simulations performed with \sacra\ using the same initial data. Although we obtain large differences in ejecta mass, with \sacra\ having more than twice the amount of ejecta than \bam, both codes predict similar values for the magnetic field. The amplification of the magnetic field during the merger is stronger with increasing resolution, as capturing KHI and MRI that trigger this amplification requires high resolution. Note that none of the simulations performed are in the convergent regime, because resolving the KHI requires an extremely high grid resolution, which is not feasible for full three-dimensional GRMHD simulations. \par 

We demonstrate that our \textit{divergence cleaning} implementation in \bam\ is able to perform reliable simulations, including the magnetic field. The scheme is simpler than other divergence-free treatments, but it does not prevent the formation of magnetic monopoles, even though this corresponds to a constraint violation near zero. However, the finite resolution error associated with other variables is larger than the errors due to divergence cleaning in all cases considered here, and all these errors should converge to zero. Our results show that uncertainties stemming from different methods for the fluxes, shock capturing methods, or Riemann solvers are more severe. 

\section*{Acknowledgements}

We thank I.~Markin for implementing the \textsc{fuka} data reader in the \bam\ code and F.~M.~Fabbri for initiating the GRMHD implementation. KK and MS also thank K.~Van Aelst for the \textsc{fuka} data reader in the \sacra.
TD and AN acknowledge support from the Deutsche Forschungsgemeinschaft, DFG, project number DI 2553/7. Furthermore, TD acknowledges funding from the EU Horizon under ERC Starting Grant, no.\ SMArt-101076369.
The simulations with \bam\ were performed on the national supercomputer HPE Apollo Hawk at the High Performance Computing (HPC) Center Stuttgart (HLRS) under the grant number GWanalysis/44189, on the GCS Supercomputer SuperMUC\_NG at the Leibniz Supercomputing Centre (LRZ) [project pn29ba], and on the HPC systems Lise/Emmy of the North German Supercomputing Alliance (HLRN) [project bbp00049]. This work is in part supported by the Grant-in-Aid for Scientific Research (grant Nos. 23H01172, 23K25869 and 23H04900) of Japan MEXT/JSPS.

\appendix

\section{Comparison General Relativistic Hydrodynamic Simulations}
\label{app:hydro}

\begin{figure}[htp]
    \centering
    \includegraphics[width=\linewidth]{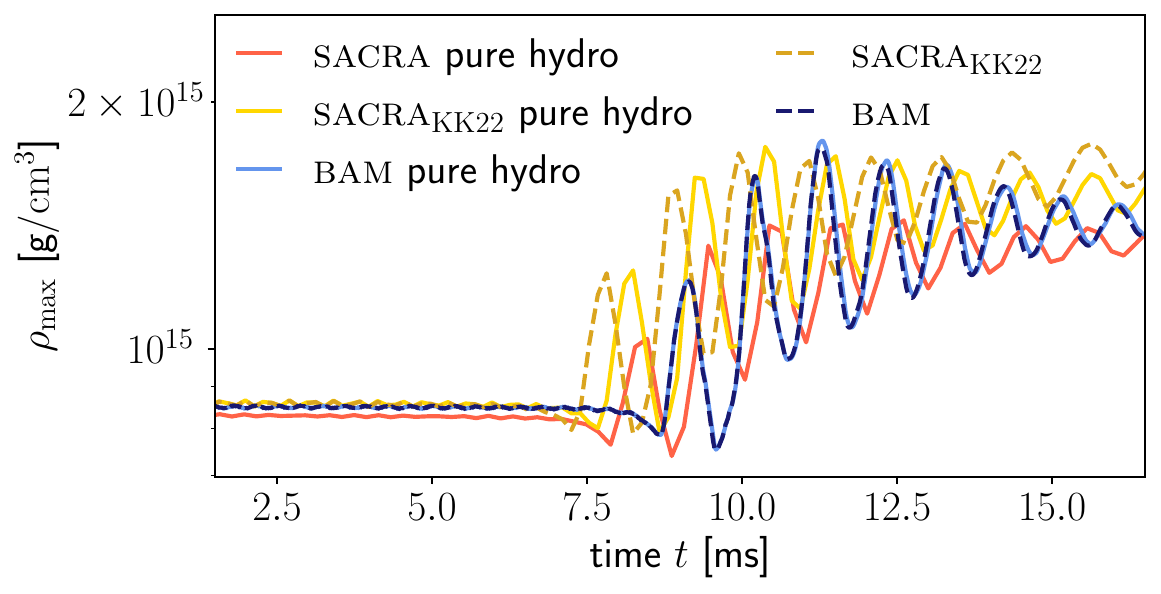}
    \caption{Comparison of the central rest-mass density in BNS simulations with \bam, \sacra, and \textsc{sacra} for resolutions R1 without magnetic field. Additionally, results for \bam\ and \sacra\ are shown for simulations with magnetic field in dashed lines. For \bam, the central rest-mass density is extracted from refinement level $l=6$.}
    \label{fig:BNShydro}
\end{figure}

As benchmark, we run the BNS simulation with R1 once without magnetic field. From these simulations, we analyze the different merger times between \bam\ and \sacra. Additionally, we run the same set-up with the \textsc{sacra} code \cite{Yamamoto:2008js}. On the one hand, \sacra\ implements an original Shibata-Nakamura version of the BSSN formulation~\cite{Shibata:1995we} and employs the gauge condition based on the Nakamura variable $F_i\equiv\delta^{jk}\tilde{\gamma}_{ij,k}$~\cite{Shibata:2003}. On the other hand, the \textsc{sacra} code implements the BSSN formulation employing the gauge conditions in a very similar way as \bam. Also, the AMR implementation based on the box-in-box in \textsc{sacra} follows \bam\ \cite{Bruegmann:2006ulg}. Therefore, in the \textsc{sacra} run, we employ the same grid configuration as \bam. In the \sacra\ run, the grid configuration is a nested grid as described in the main text. \par

We show the time evolution of the central rest-mass density in Fig.~\ref{fig:BNShydro}. The results from the corresponding simulations with magnetic fields are added as dashed lines. The different gauge condition changes the merger time in the \sacra\ simulation and indeed the differences become smaller. Since the merger time is a gauge-dependent quantity, a fair comparison between our different codes with different settings is difficult. Therefore, we focus our comparison in Sec.~\ref{subsec:BNScomparison} on gauge-independent quantities such as the ejecta mass, magnetic energy, and magnetic field strength.

\bibliography{ref.bib}

\end{document}